\documentclass[manuscript]{aastex61}

\received{June 12, 2022}
\revised{August 22, 2022}
\accepted{September 1, 2022}
\submitjournal{PASP}

\shorttitle{The Symbiotic Nova V1835~Aql}
\shortauthors{Caddy \textit{et al.}}

\begin{document}

\title{Optical Time-Series Photometry of the Symbiotic Nova V1835~Aquilae}

\correspondingauthor{Andrew Layden}
\email{laydena@bgsu.edu}

\author[0000-0002-4475-3181]{Robert V. Caddy}
\affil{University of Pittsburgh\\
100 Allen Hall, Department of Physics \& Astronomy \\
Pittsburgh, PA 15261, USA}
\altaffiliation{Bowling Green State University\\
104 Overman Hall, Physics \& Astronomy Department\\
Bowling Green, OH 43403, USA}

\author[0000-0002-6345-5171]{ Andrew C. Layden}
\affiliation{Bowling Green State University\\
104 Overman Hall, Physics \& Astronomy Department\\
Bowling Green, OH 43403, USA}

\author[0000-0002-5060-3673]{ Daniel E. Reichart}
\affil{University of North Carolina\\
Phillips Hall, CB \#3255\\
Chapel Hill, NC 27599, USA}

\author{ Joshua B. Haislip}
\affil{University of North Carolina\\
Phillips Hall, CB \#3255\\
Chapel Hill, NC 27599, USA}

\author{ Vladimir V. Kouprianov}
\affil{University of North Carolina\\
Phillips Hall, CB \#3255\\
Chapel Hill, NC 27599, USA}

\author{ Kevin M. Ivarsen}
\affil{University of North Carolina\\
Phillips Hall, CB \#3255\\
Chapel Hill, NC 27599, USA}

\author{ Justin P. Moore}
\affil{University of North Carolina\\
Phillips Hall, CB \#3255\\
Chapel Hill, NC 27599, USA}

\author{ Aaron P. LaCluyze}
\affil{University of North Carolina\\
Phillips Hall, CB \#3255\\
Chapel Hill, NC 27599, USA}

\author{ Tyler R. Linder}
\affil{University of North Carolina\\
Phillips Hall, CB \#3255\\
Chapel Hill, NC 27599, USA}

\author{ Melissa C. Nysewander }
\affil{University of North Carolina\\
Phillips Hall, CB \#3255\\
Chapel Hill, NC 27599, USA}

\begin{abstract}

We present time-series CCD photometry in the $BVRI$ passbands of the recently identified symbiotic nova V1835~Aquilae (NSV~11749) over an interval of 5.1~years with 7-14~day cadence, observed during its quiescence. We find slow light variations with a range of $\sim$0.9~mag in $V$ and $\sim$0.3~mag in $I$. Analysis of these data show strong periodicity at $419 \pm 10$~days, which we interpret to be the system's orbital period. A dip in the otherwise-sinusoidal phased light curve suggests a weak ellipsoidal effect due to tidal distortion of the giant star, which in turn opens the possibility that V1835~Aql transfers some of its mass to the hot component via Roche lobe overflow rather than via a stellar wind. We also find evidence that V1835~Aql is an S-type symbiotic star, relatively free of circumstellar dust, and include it among the nuclear burning group of symbiotics. Finally, we provide photometry, periods, and light curve classifications for 22 variable stars in the field around V1835~Aql, about half of which are newly identified.

\end{abstract}


\section{Introduction} \label{sec_Intro}

Symbiotic stars (hereafter ``SySt'') are a rare subclass of cataclysmic variables consisting of a wide interacting binary containing a cool giant star and a very hot star surrounded by gas that is excited by the latter's short-wavelength radiation. In optical spectra, this produces the unusual combination of absorption features characteristic of cool star overlain by emission features from the hot gas \citep{miko2003a, miko2015}, an unexpected combination that prompted the term ``symbiotic.''  In many cases, the hot star is a degenerate object, a white dwarf or neutron star, while the cool star is a red giant or supergiant that may be tidally distorted by its proximity to the hot star. The cool star may emit a stellar wind, some of which interacts with the hot star to generate the excited circumstellar envelope and which accretes onto its surface.
This complex arrangement can produce several types of brightness variations which may include ellipsoidal behavior, a ``reflection effect'' due to orbital motion, and irregular variations related to accretion.   In some cases, the cool star pulsates, providing a Mira-like element to the light curve \citep{miko2003a, angeloni14}.  Different stars exhibit different degrees of these behaviors, and are collected in the inhomogeneous  ``Z Andromedae'' (ZAND) class of objects \citep{samus2009}.
The AAVSO Variable Star Index (VSX) currently lists 157 stars with this classification.\footnote{As of 2022 May, there are over two million objects in the VSX, see  \url{http://www.aavso.org/vsx/}. }

An even rarer, related group is the NC class of slow novae \citep{samus2009}. The VSX lists only six such stars. Outbursts of up to 10 mag are characterized by a slow rise to peak light, which is sustained for about a decade and followed by a very slow decline. The spectra of some members can resemble those of SySt, and long-period light variations like those in SySt have been observed during quiescence. \citet{miko2010} provides an excellent description of these symbiotic novae.

A recent addition to the short list of slow, symbiotic novae is V1835~Aquilae. Earlier identified as NSV~11749 \citep{ksdkp15}, the star was first discovered to be variable by \citet{luyten1937}.  \citet{williams2005} derived photometry from plates in the Harvard collection showing an outburst \textit{circa} 1903 that reached a photographic magnitude of 12.5 then slowly faded below the plate limit of $m_{\rm ptg} \approx 15$ mag by 1912.  It remained below this limit through the end of the plate sequence in 1988, though it had been detected on a few deeper plates at $\sim$17 mag.  Based on this outburst, Williams suggested that V1835~Aql is either a slow nova or a FU Orionis pre-main sequence star. %
\citet{bond12} presented optical and
infrared spectra of V1835~Aql showing Balmer and helium emission lines on a continuous spectrum of a M1-M2 giant. Their spectrum also showed a broad emission line at  6825~\AA~
 caused by light from quintuply-ionized oxygen (O~VI) at 1032~\AA,  Raman-scattering off of cold, neutral hydrogen, an effect only seen in SySt.
\citet{bond12} provided an excellent discussion of the preceding interpretations of V1835~Aql and gave convincing evidence that this intriguing object is a symbiotic star.

V1835~Aql was detected independently by a survey for H-alpha emission \citep{kohoutek} and the ChaMPlane project to detect cataclysmic variables \citep{grindlay2005, rogel2006}.  
\citet{wehrung13} monitored V1835~Aql photometrically in the $VI$ passbands for six months in 2012 and
saw slow variations that supported the assessment of \citet{bond12} that V1835~Aql is a symbiotic nova. %
However, the duration of that time-series was too short to determine the periodic nature of the star. \citet{wehrung13} also provided a spectrum of the star (which did not cover the SySt diagnostic line at 6825~\AA) and argued for a reddening value of $E(B-V) = 0.67 \pm 0.10$ mag. In addition, they found nine additional variable stars located nearby, many of which are now listed in the VSX under the name ``[WLR2013]~\#'' where the number corresponds to the identification shown in Table~2 of \citet{wehrung13}. 

In this paper, we report continued time-series CCD photometry of V1835~Aql aimed at 
clarifying the nature of this object in quiescence and determining its orbital period. We also update the photometric properties of the nearby variable stars and detect new ones. In Section~\ref{sec_Obs} we describe the new observations, and outine the stellar photometry in Section~\ref{sec_Reds}. We analyze the light curve of V1835~Aql in Section~\ref{sec_syst}, and discuss the light curves of the other variable stars in Section~\ref{sec_olcs}. We summarize our results in Section~\ref{sec_concl} and suggest future observations to further clarify the nature of V1835~Aql.

\section{Observations}  \label{sec_Obs}

We obtained images of V1835~Aql between 2012 and 2017 using the Panchromatic Robotic Optical Monitoring and Polarimetry Telescopes (PROMPT) located on Cerro Tololo in Chile \citep{reichart05}.  During this time, we used a variety of telescopes and CCD cameras, as summarized in Table~\ref{tab_telequ}, in order to maintain a regular monitoring cadence as the availability of the PROMPT equipment shifted over time; a secondary goal was to minimize equipment changes in order to maximize the internal consistency of our data set.  Throughout this time, images were acquired using the Johnson $V$ filter and the Cousins $I$ filter.\footnote{These include the PROMPT observations taken from July to November 2012 and presented in \citet{wehrung13}.  We have reanalyzed these images herein to maximize consistency with the PROMPT images taken after 2012, as described below.}   Additional images were obtained contemporaneously with the $VI$ images in the Johnson $B$ filter between 2013-03-09 and 2014-09-26 (on a total of 51 nights), and in the Cousins $R$ filter from 2012-08-05 to 2013-11-05 (40 nights).  On most nights, three images were obtained in each filter, with the telescope position offset by a few pixels between each exposure.  Observing nights were separated by 7-14 days in order to maintain a slow, regular cadence suited to the long periods of the variable stars found by \citet{wehrung13}.  All images were processed using the standard methods of bias, dark-current, and flat-field correction.

\begin{deluxetable}{cccrccr}
\tablecaption{Telescopes and Instruments\label{tab_telequ}}
\tablehead{
\colhead{Dates} & \colhead{Telescope} & \colhead{Camera} & \colhead{Scale} & \colhead{Field} &
\colhead{Filters} & \colhead{Nights}  
}
\startdata
2012-05-16 to 2012-07-11 & BGSU~0.5-m & Ap6e  & 1.2 & $21 \times 21$ & $VI$ & 20 \\
2012-07-24 to 2014-11-04 & PROMPT5 0.4-m & P5AA & 0.6 & $10 \times 10$ & $BVRI$ & 78 \\
2014-07-29 &            PROMPT5 0.4-m & P5AU & 0.6 & $10 \times 10$ & $BVI$ & 1 \\
2015-03-02 to 2016-05-20 & PROMPT5 0.4-m & P5AU & 0.6 & $10 \times 10$ & $VI$ & 36  \\
2016-06-23 to 2016-11-05 & PROMPT1 0.6-m & P1AU & 1.2 & $24 \times 24$ & $VI$ & 14 \\
2017-02-28 to 2017-06-29 & PROMPT5 0.4-m & P5AU & 0.6 & $10 \times 10$ & $VI$ & 8  \\
\enddata
\tablecomments{The image scale and field of view are given in units of arcsec~pix$^{-1}$ and arcminutes, respectively, in columns 4 and 5.}
\end{deluxetable}

When combined with the early observations made with the 0.5-m telescope at Bowling Green State University (BGSU; see Table \ref{tab_telequ}) described in \citet{wehrung13}\footnote{These data were not reanalyzed herein; we use the BG photometry listed in \citet{wehrung13}.}, we have photometry from a total of 1190 images taken on 157 independent nights spanning the Julian dates of 2,456,064 to 2,457,933 days (5.1 years).   The median seeing of the $V$-band images was 2.5 arcsec full-width at half-maximum, and 2.7 arcsec for the $I$-band images, though in all filter sets the seeing varied from 1.6 to 3.7 arcsec at the 10\% and 90\% points of the seeing distribution.

~
~

\section{Stellar Photometry}  \label{sec_Reds}

We analyzed all the PROMPT images 
following the methods described in \citet{wehrung13}.  We obtained a list of star positions and instrumental magnitudes for each image using the DAOPHOT~II point-spread function fitting software \citep{stetson87}.  These lists were merged using DAOMASTER and the instrumental magnitudes for each image were refined using ALLFRAME \citep{stetson94} utilizing a star list based on the best quality images.  Using a uniform star list helps to mitigate the effect of blending and merging of stellar profiles resulting from the range of seeing conditions present in our images.  We separated the images into sets based on the telescope, camera and filter used (see Table \ref{tab_telequ}) and processed them separately with DAOMASTER and ALLFRAME.  In doing so, we found that during the ``P5AU'' set, and again during the ``P1AU'' set, there was an abrupt rotation in the images (perhaps due to removal of the camera for maintenance) that complicated the DAOMASTER/ALLFRAME process; we split the images into separate pre- and post-rotation sets and found that the resulting DAOMASTER/ALLFRAME analysis ran well. 

For each telescope-camera-filter set, we used the instrumental and standard magnitudes ($m_i$ and $m_s$ respectively) of  fifteen in-field secondary standard stars established in \citet{wehrung13} to obtain the coefficients $c_0$ and $c_1$ in the transformation equation
\begin{equation}
  m_i - m_s = c_0 + c_1~(V-I)_s,
\end{equation}
where $(V-I)_s$ is the standard color.  The rms scatter around these least-squares fits were typically about 0.026 mag in $V$, about 0.044 mag in $I$, and 0.019 mag in $R$.  For $B$, we did not have standard magnitudes from \citet{wehrung13}, so we downloaded data for six uncrowded comparison stars from the APASS database \citep{henden14,henden15}\footnote{Data from APASS Data Release 9 were acquired from \url{http://www.aavso.org/apass/}. }. Though the formal uncertainty in the each APASS $B$ magnitude was relatively large, $\sim$0.06 mag, the rms scatter around the $B$ fit was only 0.031 mag, indicating a reliable calibration.

\begin{figure}
\plotone{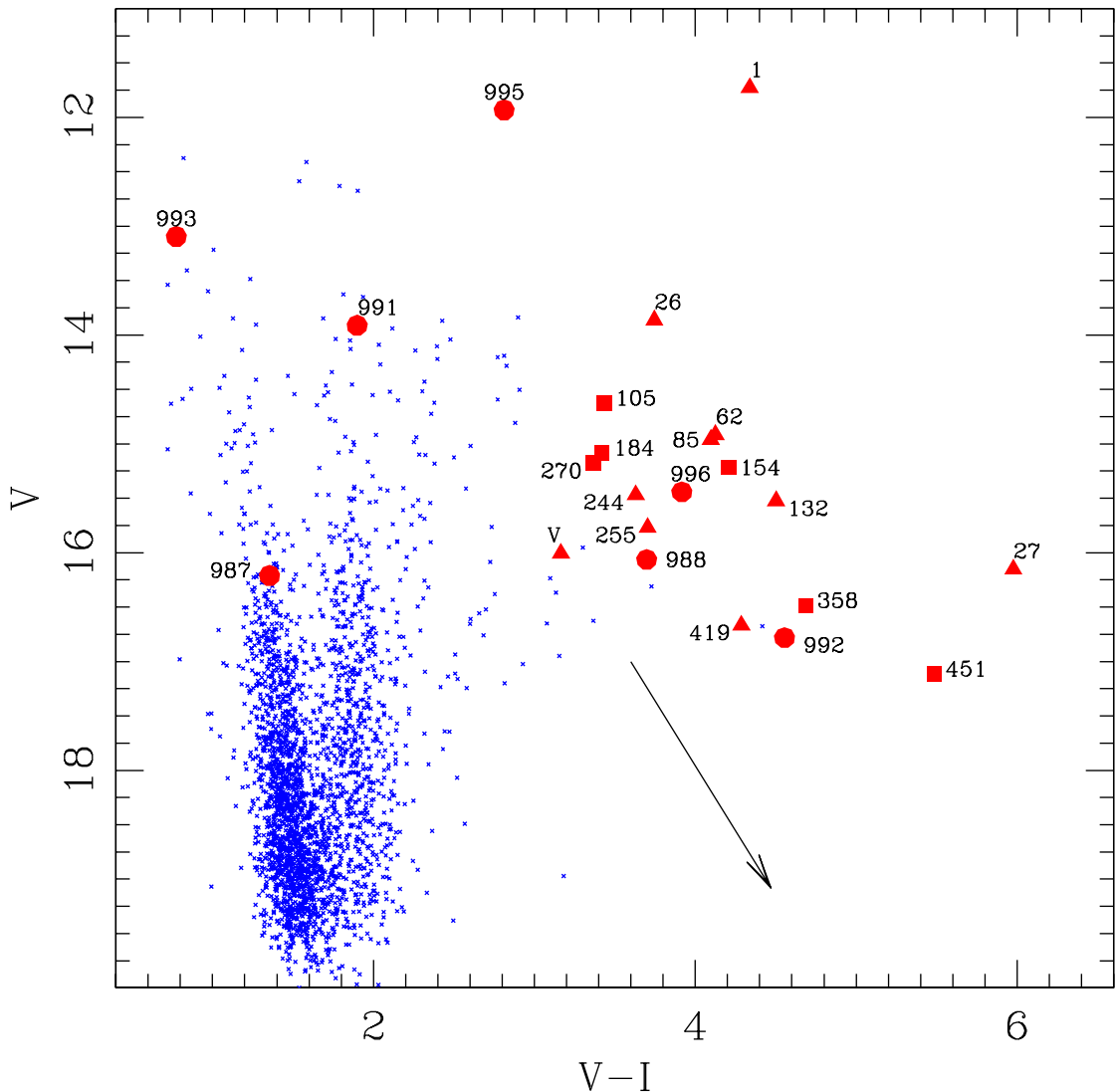}
\caption{The $VI$ color-magnitude diagram for stars in the P5AA data set.  Variable stars detected in \citet{wehrung13} are marked with triangles, along with the seven new variables found in the P5AA field (circles) and the six new variables from the P1AU field (squares).  Labels indicate the variable star identification numbers shown in Table~\ref{tab_coords} and the ``V'' marks V1835~Aql.  The diagonal arrow shows the reddening vector from \citet{wehrung13}.
  \label{fig_cmdvars}}
\end{figure}

We then applied the coefficients to every star in the set to obtain the average standard magnitude of each star in each passband.  The resulting color-magnitude diagram for the P5AA set is shown in Figure~\ref{fig_cmdvars}.  We estimate the overall zero-point uncertainty in the photometric calibrations to be approximately 0.02 mag in $V$ and $I$, and about 0.04 mag in $R$ and $B$.  The lower resolution of the APASS images and consequent potential for blending of faint neighbor stars with each comparison star leads to a concern that a systematic error in our $B$ magnitude transformation could be present.

To detect variable stars, we used the \citet{welch-stetson93} method on our paired $V$ and $ I$ magnitudes from the 2.3-year P5AA data set to search for correlated brightness changes.  We found 17 variable stars, ten of which had been identified by \citet{wehrung13}, including V1835~Aql.  We also applied the  \citet{welch-stetson93} analysis on the P1AU data set; though the data span a much shorter interval of 0.4 years, the much larger field of view could contain many variable stars in addition to the three originally detected by \citet{wehrung13} from the BGSU data set (\#85, \#132 and \#419).  
We again used the  \citet{welch-stetson93} variability index to identify variable candidates in the P1AU set, but adjusted our selection criteria from those used in the P5AA search to focus on stars that show slow, correlated variations in $V$ and $I$ in their time-magnitude plots, as well as red mean colors. No doubt this has biased our search toward finding long period variable stars in the P1AU fields, within which we detected six new variable stars.

The names and equatorial coordinates of these 23 variables, along with their $J$ and $K_s$ photometry from the Two Micron All-Sky Survey's Point Source Catalog \citep{skrutskie06}, are listed in Table~\ref{tab_coords}. The uncertainty in the $J$ and $K$ magnitudes for most stars is between 0.02 and 0.04 mag, the exception being the bright star \#1 with $J_{err} = 0.22$ and $K_{err} = 0.29$ mag. The 2MASS photometry confirms that most of the new variable stars are quite red. 

Also shown in Table~\ref{tab_coords} is the distance to each star in kiloparsecs, computed from the trigonometric parallax listed in the early third data release ``EDR3" \citep{GaiaEDR3} of the Gaia space astrometry mission \citep{GaiaSat}.
Cases in which the parallax error was more than half the parallax value are of little direct use, and are excluded from the table. The exception is V1835~Aql, for which we find a photometric distance of 8-11~kpc as described at then end of Sec.~\ref{sec_syst}.

\begin{deluxetable}{rcccccl}
\tablecaption{Variable Star Coordinates and Literature Data \label{tab_coords}}
\tablehead{
\colhead{ID\#} & \colhead{RA (J2000)} & \colhead{Dec (J2000)} & \colhead{$J$} & \colhead{$K_{\rm{s}}$} & \colhead{$d_{\rm{kpc}}$} & \colhead{Comment} 
}
\startdata
417  & 19:07:42.4 & +00:02:51 & 10.794 & 9.351 &  $31^{+147}_{-14}$ & V1835~Aql, $d_{\rm{phot}}$ = 8-11 kpc \\
  1  &  19:07:44.3 & +00:07:10 &  4.916 & 3.308 & $0.88^{+0.05}_{-0.04}$ & ASAS~190744+0007.1 \\
 26  & 19:07:55.1 & +00:05:30 & 7.939 & 6.489 & $3.9^{+0.9}_{-0.6}$ & -- \\
 27  & 19:07:37.5 & +00:06:09 & 6.159 & 4.301 & $3.2^{+1.9}_{-0.9}$ & IRAS~19050+0001\\
 62  & 19:07:42.4 & +00:02:22 & 8.397 & 6.920 & $5.7^{+2.1}_{-1.2}$ & -- \\
 85  & 19:07:30.2 &--00:02:51 & 8.470 & 6.965 & $7.9^{+5.2}_{-2.2}$ & BGSU + P1 fields only \\
132  & 19:07:10.4 &--00:01:55 & 8.379 & 6.841 & \nodata & BGSU + P1 fields only \\
244  & 19:07:25.1 & +00:00:59 & 9.774 & 8.302 & $17^{+21}_{-6}$ & -- \\
255  & 19:07:48.5 & +00:05:32 & 9.964 & 8.460 & $24^{+106}_{-11}$ & -- \\
419  & 19:07:17.2 & +00:00:40 & 10.302 & 8.735 & \nodata & BGSU + P1 fields only \\
991 & 19:07:48.7 & --00:01:07 & 10.689 & 9.688 & $2.78^{+0.20}_{-0.18}$ & new \\
992 & 19:07:45.0  & +00:04:05 &  9.578 & 8.006 & \nodata & new \\ 
993 & 19:07:55.3 & --00:01:19 & 11.698 & 11.346 & $2.22^{+0.10}_{-0.10}$ & new \\ 
995 & 19:07:26.7 & +00:03:54 & 7.313 & 5.968 & $2.32^{+0.19}_{-0.17}$ & new \\
996 & 19:07:42.8 & +00:04:41 & 9.250 & 7.728 & $7.3^{+2.7}_{-1.5}$ & new \\
987 & 19:07:22.5 & +00:06:54 & 14.211 & 13.714 & $1.49^{+0.10}_{-0.09}$ & new \\
988 & 19:07:23.3 & +00:07:17 & 10.236 & 8.736 & $15^{+17}_{-5}$ & new \\
105 & 19:08:13.7 & +00:00:19 &  9.138 & 7.670 & $5.8^{+1.4}_{-0.9}$ & new P1+BG \\      
154 & 19:08:06.5 & --00:03:13 &  8.637 & 7.199 & $4.4^{+1.8}_{-1.0}$ & new P1+BG \\
184 & 19:07:11.2 & +00:02:27 &  9.565 & 8.122 & $16^{+14}_{-5}$ & new P1+BG \\    
270 & 19:07:26.8 & --00:03:47 &  9.947 & 8.565 & $6.2^{+1.1}_{-0.8}$ & new P1+BG \\ %
358 & 19:07:36.8 & +00:14:03 &  8.963 & 7.476 & $9^{+13}_{-3}$ & new P1+BG \\
451 & 19:07:43.2 & +00:08:24 &  7.828 & 6.133 & $10^{+80}_{-5}$ & new P1+BG \\
\enddata
\end{deluxetable}

For each of these 23 variable stars plus the 15 comparison stars discussed above, we extracted the instrumental time-series photometry from ALLFRAME, and combined them differentially to obtain standard $BVR I$ magnitudes.  For a particular variable star on a specific image, we calculated
\begin{equation}
  m_{s,v} = m_{s,c} + m_{i,v} - m_{i,c} - c_1~ [ (V-I)_{s,v} - (V-I)_{s,c} ],
\end{equation}
where the $v$ and $c$ subscripts refer to the variable and comparison star, respectively.  We repeated this calculation for each of the $N_{\rm{c}}$ comparison stars visible on that image to arrive at $N_{\rm{c}}$ separate estimates of the variable star's magnitude, then took the median of the $N_{\rm{c}}$ estimates as the best estimate.  We also computed the standard deviation of the mean ($\sigma_{\rm{m}}$) from the $N_{\rm{c}}$ estimates as the best estimate of the uncertainty in the variable star's standard magnitude.  Another estimate of this uncertainty, $\epsilon$ was produced by propagating the photometric errors from ALLFRAME and the uncertainty of the comparison stars' standard magnitudes through Eqn.~2.  These values are listed in Table~\ref{tab_timeser}, along with the heliocentric Julian date (HJD) of the observation, the FWHM seeing (in arcsec), the airmass at the time of observation, and an integer on a 1-5 scale describing the visual quality of stellar profiles on the image ($Q$ = 1 indicates round star with a radial profile plot having little scatter; $Q$ = 5 marks a very elongated profile due to tracking errors and/or an out of focus stellar profile, both of which have large scatter in their radial profile plots).  The photometry set is also indicated.

\begin{deluxetable}{lrrrrrcccll}
\tablecaption{Time-Series Photometry of the Variable Stars \label{tab_timeser}}
\tablehead{
\colhead{Star} & \colhead{HJD} & \colhead{Mag} & \colhead{$\epsilon$} & 
\colhead{$\sigma_{\rm{m}}$} & \colhead{$N_{\rm{c}}$} & \colhead{Filter\tablenotemark{a}}  & \colhead{FWHM} & 
\colhead{Airmass} & \colhead{$Q$\tablenotemark{b}} & \colhead{Set}  
}
\startdata
417 & 2456064.7500 & 16.206 & 0.075 & 0.005 & 15 & 1 & 3.3 & 1.88 & 2 & BGSU \\
417 & 2456064.7510 & 12.814 & 0.017 & 0.010 & 15 & 2 & 3.3 & 1.88 & 2 & BGSU \\
417 & 2456064.7549 & 16.043 & 0.065 & 0.005 & 15 & 1 & 3.4 & 1.83 & 2 & BGSU \\
417 & 2456064.7559 & 12.823 & 0.013 & 0.009 & 15 & 2 & 3.4 & 1.83 & 2 & BGSU \\
\nodata & \nodata & \nodata & \nodata & \nodata & \nodata & \nodata & \nodata & \nodata & \nodata & \nodata \\
001 & 2456149.5079 & 11.705 & 0.020 & 0.021 &  6 & 1 & 4.4 & 1.43 & 3 &  P5AA \\
001 & 2456149.5476 &  9.698 & 0.011 & 0.011  & 5 & 3 & 3.8 & 1.24 & 3 &  P5AA \\
001 & 2456149.5480 &  7.304 & 0.049 & 0.030  & 5 & 2 & 1.9 & 1.24 & 2 &  P5AA \\
\nodata & \nodata & \nodata & \nodata & \nodata & \nodata & \nodata & \nodata & \nodata & \nodata & \nodata \\
451 & 2457700.5244 & 11.563 & 0.038 & 0.012 & 15 & 2 & 3.1 & 2.22 & 2 & P1AU \\
451 & 2457700.5247 & 11.551 & 0.033 & 0.013 & 15 & 2 & 3.0 & 2.23 & 2 & P1AU \\
 \enddata

\tablecomments{Table \ref{tab_timeser} is published in its entirety in the electronic edition; a portion is shown here for guidance regarding its form and content.}
\tablenotetext{a}{An integer code is used to describe the filter employed: 1 = $V$, 2 = $ I$, 3 = $R$ and 4 = $B$.}
\tablenotetext{b}{The image quality index ``$Q$'' is described in Section~\ref{sec_Reds}.}
\end{deluxetable}

The locations of the 23 variable stars are marked in the color-magnitude diagram of Figure~\ref{fig_cmdvars}.  We note that many of the variable stars are significantly more red than the comparison stars ($0.6 < V-I < 2.3$ mag), so that uncertainty in establishing the coefficient $c_1$ in the equation above can lead to a systematic zero-point error in the resulting magnitude of each variable star.

\section{The Light Curve of V1835~Aql}  \label{sec_syst}

\begin{figure}
\plotone{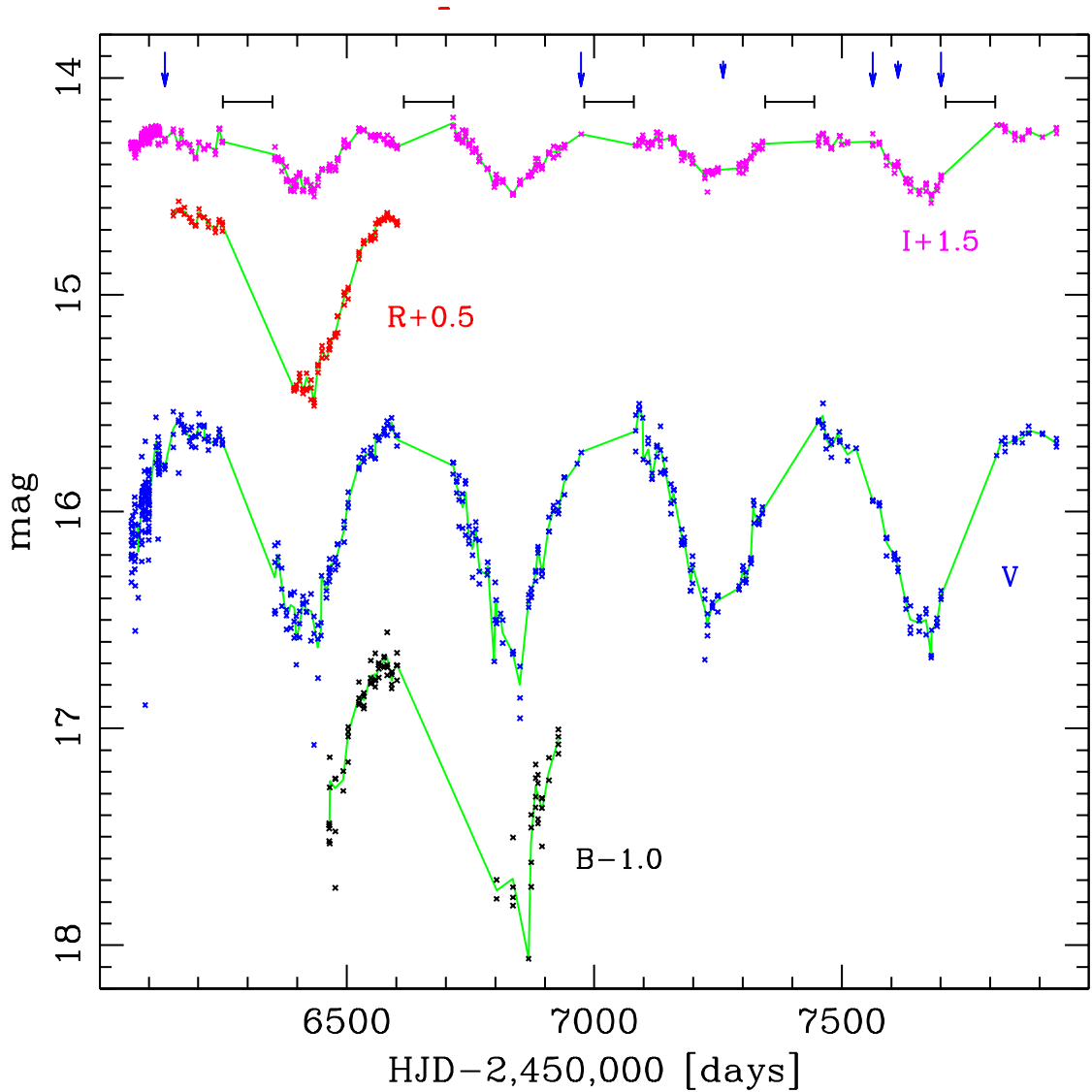}
\caption{The $BVRI$ photometry of V1835~Aql is shown as a function of Julian date, shifted vertically as indicated.  The crosses indicate magnitudes derived from individual images, and the green lines connect nightly-mean points.  The bars along the top 
indicate when the field was behind the Sun and therefore unobservable.  
The long vertical arrows 
mark the transition times between the different telescope-instrument sets from Table~\ref{tab_telequ}; from left to right, the intervals are BGSU, P5AA, P5AU, P1AU and P5AU.  The shorter arrows indicate transitions between the image rotation sets described in Sec.~\ref{sec_Reds}.
  \label{fig_lc_syst}}
\end{figure}

The light curve of the symbiotic star V1835~Aql is shown in Figure \ref{fig_lc_syst}.
For images obtained within a 12-hour span, the resulting photometry was combined
using an error-weighted mean to obtain a single data point per filter per night, shown connected by the green line.  The 
transition times between the different telescope-camera sets are indicated.  The photometry appears to be smooth and continuous throughout the time series, indicating that we successfully transformed the different sets of instrumental photometry onto a consistent standard system.  Additional tests on several non-variable check stars confirmed that there were no significant jumps at the transitions between data sets that might indicate problems with the photometric calibrations.
The $\sim$100-day gaps in the time series resulted because the star was below an elevation of $\sim$20 degrees all night between mid-November and late-February each year.

Despite the observational scatter, Figure \ref{fig_lc_syst} shows fairly regular cyclic behavior with an amplitude of $\sim$0.9 mag in $V$ and $\sim$0.3 mag in $I$.  Several points are worth noting.  First, the level of maximum light appears to be similar from peak to peak in both filters, though the presence of the data gaps makes this difficult to ascertain.  Second, the level of the minima 
is similar from cycle to cycle, though the star is faint in $V$ at minimum light and the resulting scatter is large.  Third, the light curve shape is consistent with that of a contact binary (W~Ursae Majoris type variable, EW) or ellipsoidal variable (ELL) with a period of $\sim$850 days, or with a pulsating variable with a period half that length.

Analysis of the $V$-band data using the discrete Fourier transform method implemented in \textsc{VStar} \citet{benn12} yielded a period 423 of days.  A similar application using the $I$-band data resulted in a period of 431 d, but the lower amplitude in this passband yielded a peak with lower power. We obtained another estimate of the period by estimating the times of minimum light seen in Figure~\ref{fig_lc_syst} and fitting them with a linear ephemeris of the form $JD_{min} = P~N_{cyc} + E_0$, to obtain $P = 419 \pm 10$~d and $E_0 = 2,456,834 \pm 10$~d, which we adopt as the best description of the single peak-trough cycle seen in Figure~\ref{fig_lc_syst}.

\begin{figure}
\plotone{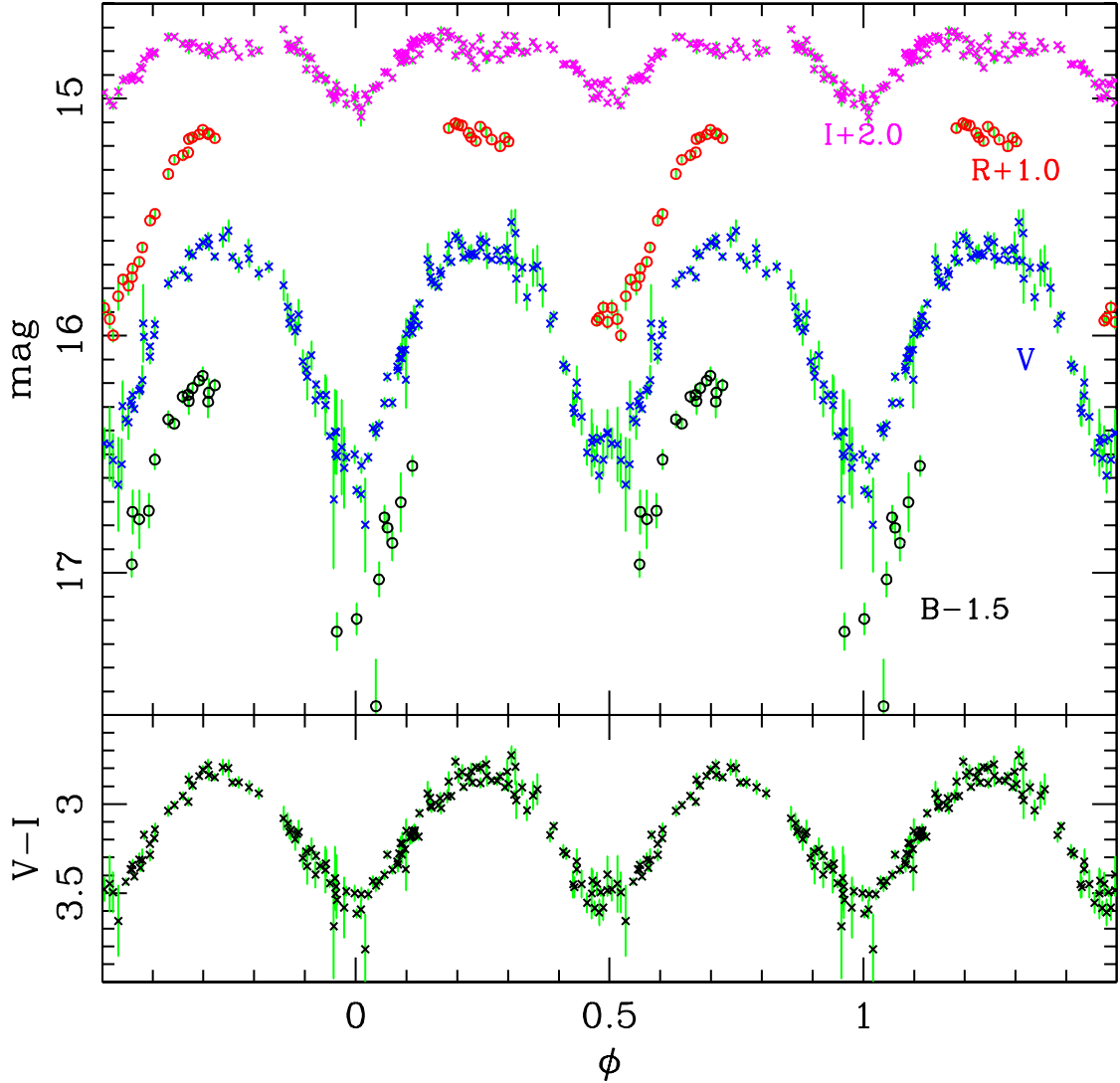}
\caption{The $BVRI$ light curves (upper panel) and $V-I$ color curve (lower panel) of V1835~Aql phased with a period of 838~days and time of minimum light $E_0 = 2,456,834$~d. These nightly-averaged data were shifted vertically by the indicated amounts for convenience of display. 
  \label{fig_lc_phi1}}
\end{figure}

Figure \ref{fig_lc_phi1} shows the light and $V-I$ color curves of V1835~Aql phased with twice this period, the expected orbital period if the system is expressing ellipsoidal variation similar to that seen in a sample of ELL-type stars in the Large Magellanic Cloud (LMC) by \citet{soszynski04}.  In our phased light curves, 
the primary eclipse at $\phi = 0$ is marginally deeper ($\sim$1.5-$\sigma$) than the secondary eclipse.  This slight difference is common among ellipsoidal variables with a main-sequence secondary star \citep{soszynski04} and has been attributed to limb-darkening and gravity-darkening in the elongated ``tip'' of the primary star that points toward the secondary, so that the deeper light curve minimum occurs when this tip is pointing toward the observer; the difference tends to be largest when the primary fills its Roche lobe \citep{hall90b}.

However, the light curve amplitude of V1835~Aql, $\sim$0.3~mag in $I$, is large compared with those of ellipsoidal variables in the LCM having periods longer than 500~days \citep{soszynski04}, and the amplitude in $V$ is larger still,  $\sim$0.9~mag.  \citet{sj96} noted that $V$-band amplitudes of classical ellipsoidal variables rarely exceed 0.1~mag, with an upper limit of $\sim$0.4~mag, and their sample light curves show that photometric amplitudes of ellipsoidals are roughly the same size at all visible wavelengths.  The differing amplitudes across $VRI$ observed in  V1835~Aql can not be explained by simple variations in effective surface area due to viewing a tidally-distorted star at varying orbital phases \citep{leahy15}.   Similarly, the sinusoidally-varying $V-I$ color of V1835~Aql shown in the lower panel of Figure \ref{fig_lc_phi1} is difficult to explain in a tidally distorted star as described above.  We would expect the color to be reddest when the primary star's cool tidal tip is pointed toward us during the deeper light curve minimum, and bluest when it is pointing away during the shallower minimum.  Together, the size of the amplitudes and the timing of the color-curve leads us to infer that V1835~Aql is not a simple ellipsoidal variable with an orbital period of 838~d.

The sinusoidal color variations are more in-keeping with a SySt system in which a hot, compact component with a surrounding ionized nebula orbits a cool red giant star.  One color cycle would thus correspond to one orbital cycle of $P_{orb} = 419$~d, with the object appearing brightest and bluest when the hot nebula is visible; the object would gradually appear fainter and redder as the nebula rotates out of view behind the cool red giant.

\begin{figure}
\plotone{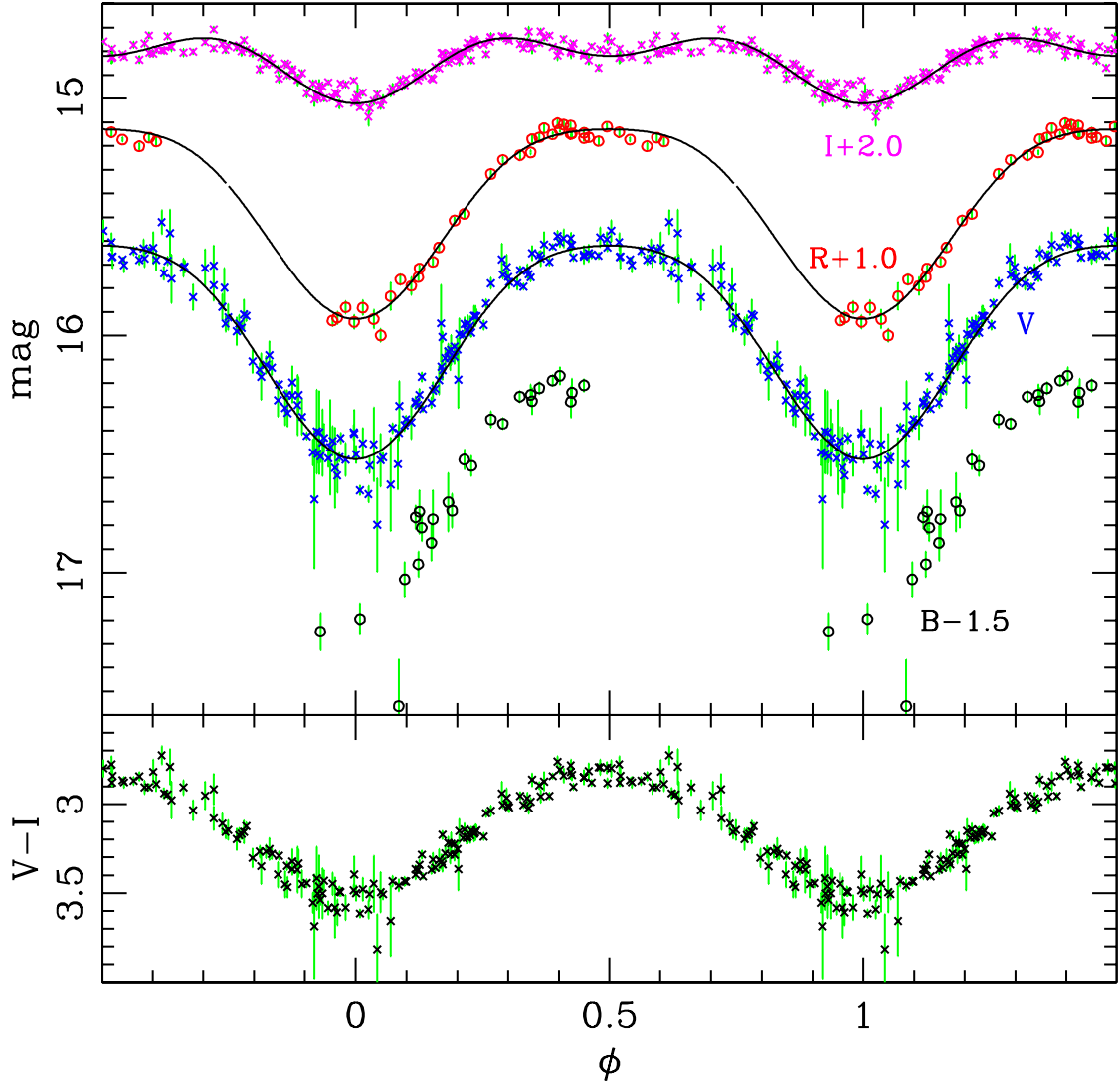}
\caption{The light and color curves of V1835~Aql, analogous to those shown in Figure~\ref{fig_lc_phi1} but phased with a period of 419~d and $E_0 = 2,456,834$~d.  The black curves are models described in Section~\ref{sec_syst}.
  \label{fig_lc_phi2}}
\end{figure}

Figure~\ref{fig_lc_phi2} shows the star's light and color curves phased with the  419-d period.  No sharp-edged eclipses like those seen in Figures~9 and 10 of \citet{grom13} are evident. This suggests that the orbital inclination is too small 
for the compact object to pass behind the red giant, or that the luminosity of the compact object relative to the red giant is too low to produce an eclipse (or both).  However, the inclination is sufficient for the red giant to partially obscure the luminous nebula during the orbital cycle, thus causing the principal modulation in the light and color curves.

Evidence of a very shallow dimming episode is visible at $\phi = 0.5$; it appears as a dip in $I$ but only flattens the top of the $V$ and $R$ curves.  Modeling the light curves with \textsc{VStar} suggests a sinusoid with $P_{orb}/2$ and an amplitude of 0.08~mag matches the light curves in $VRI$ (as shown by the solid curves in Figure~\ref{fig_lc_phi2}), leaving the $V-I$ color as an unaffected sinusoid. 
This behavior is very similar to that seen in the multi-band light curves of YY~Her \citep{miko02yy} and LT~Del \citep{munari21}, in which secondary dips of $\sim$0.1~mag are seen at $\phi = 0.5$ in the $VRI$ passbands; their overall range of variation increases in passbands at shorter wavelengths, so that the dip merely flattens the peak of the underlying sinusoid in $B$ and (for YY~Her) does not affect it perceptibly in $U$. In the photometry of another system, T~CrB, the secondary minimum is deeper and 
is clearly visible in $B$ as well as in $VRI$ \citep{munari21}, whereas in our photometry of V1835~Aql, the dip is shallower and is clearly visible only in $I$. In summary, the dips are seen in the redder passbands for all these objects, but the extent to which dips remain visible at shorter wavelengths varies from one object to the next.

Such secondary dips in SySt are widely attributed to ellipsoidal variations caused by the asymmetrical tidal distortion of the red giant star (see \citet{miko02yy}, \citet{munari21} and references therein). Because of their primarily geometric cause, the depth of the ellipsoidal dips is roughly constant with wavelength in a given star, as shown for classical ellipsoidal variables in \citet{sj96}. Meanwhile, the principle sinusoidal behavior, caused by varying visibility of the hot nebular material over the course of an orbit, tends to be stronger in shorter wavelengths, and thus dominates the ellipsoidal modulation in the blue.  The degree to which the irradiated nebula or the red giant photosphere dominates the light curve in a given passband, say $V$, likely depends on the temperature and luminosity of the nebular material, which in turn depends on the luminosity of the hot, compact stellar component that irradiates it. \citet{munari21} and \citet{grom13} show $V$-band light curves of a number of SySt that express a range of primary (nebular) and secondary (ellipsoidal) modulation, which indicates a range of luminosities in the compact stars across these systems. Indeed, the systems with stronger nebular emission and larger blue amplitudes have been associated with white dwarfs that fuse the material accreted from the red giant continuously, as soon as it reaches the surface (nuclear burning-type, \textit{burn}-SySt), whereas the white dwarfs in the systems where the ellipsoidal modulation of the red giant dominates are thought to be powered solely by the gravitational energy of accretion (accreting-only, or \textit{accr}-SySt), according to \citet{munari21} and references therein. Because of the strong nebular modulation seen in V1835~Aql, we place this system among the more commonly-observed \textit{burn}-SySt subclass. It is intriguing that this system also produced a strong nova-like outburst observed in 1903 \citep{williams2005}.

Because the secondary, ellipsoidal dip should be more evident at longer wavelengths, we advocate for new time-series observations of V1835~Aql in the near infrared. Also, obtaining $UB$ photometry over the full orbital cycle would verify our finding that the ellipsoidal modulation is dominated by the nebular emission at theses energies, and enable a more quantitative flux model of the type performed by \citet{miko02yy} on YY~Her.  In addition, radial velocities obtained at different orbital phases would help to confirm the 419-d orbital period rather than twice this (ELL variable), and could also lead to mass estimates for the two stars as shown by \citet{miko2003a}.

The variations of V1835~Aql appear quite regular from cycle to cycle, indicating that variability of the red giant due to chromospheric (starspot) activity or pulsational Mira-like behavior, often seen in symbiotic variables (e.g., \citet{angeloni14}), is small, $\lesssim$0.1 mag. This is consistent with V1835~Aql being an S-type (stellar), rather than the rarer D-type (dusty), symbiotic system \citep{miko2003a}. The infrared colors of V1835~Aql listed in Table~\ref{tab_coords}, when corrected for interstellar reddening, are also in good agreement with those of normal M3-4 giants \citep{kuc05}, indicating that we are viewing the photosphere of the red giant in V1835~Aql rather than its circumstellar dust, and thus it is a SySt of S-type.

We determined the median $BVRI$ magnitudes and light curve extrema from Figure~\ref{fig_lc_phi2}, and report them in Table~\ref{tab_lcparams}, where the brightest and faintest 2\% of the data points were eliminated to reduce the effects of outliers and photometric scatter. The column labeled $R_f$ shows the photometric range in filter $f$ as the difference between the faintest and brightest magnitudes listed.

The parallax-based distance for V1835~Aql shown in Table~\ref{tab_coords} has large uncertainties, indicating that this star remains outside the effective range of Gaia trigonometric parallaxes as of EDR3, though this situation may improve as future data releases become available.  We can estimate the distance to V1835~Aql from the $V$ magnitudes in Table~\ref{tab_lcparams}, the reddening $E(B-V) = 0.67$ mag \citep{wehrung13}, and an absolute magnitude of $-0.7$ mag based on the M3-4 spectral type \citep{wehrung13}.  While the uncertainties in these values are large, we estimate that V1835~Aql lies 8-11 kpc from the Sun, 5--6 kpc from the Galactic center and about 0.6 kpc below the Galactic plane. We also note that an upper limit on the distance can be obtained by assuming the red giant fills its Roche lobe; following the arguments in \citet{miko02yy} for YY~Her, one finds that V1835~Aql must lie within $\sim$15 kpc of the Sun.

\begin{deluxetable}{lrrrrr}
\tablecaption{Photometric Properties of V1835~Aql \label{tab_lcparams}}
\tablehead{
\colhead{Filter} & \colhead{Median} & \colhead{Bright} & \colhead{Faint} &  \colhead{$R_f$}  & \colhead{$N_{obs}$}  
}
\startdata
$V$ &  16.007 & 15.55 & 16.58 & 1.03 & 335  \\
$I$ &  12.842 & 12.72 & 13.03 & 0.31 & 332  \\
$R$ & 14.257 & 14.10 & 14.99 & 0.89 &  99  \\
$B$ & 18.027 & 17.65 & 18.79 & 1.14 & 78  \\
\enddata
\end{deluxetable}

\section{The Light Curves of other Variable Stars}  \label{sec_olcs}

\begin{figure}
\plotone{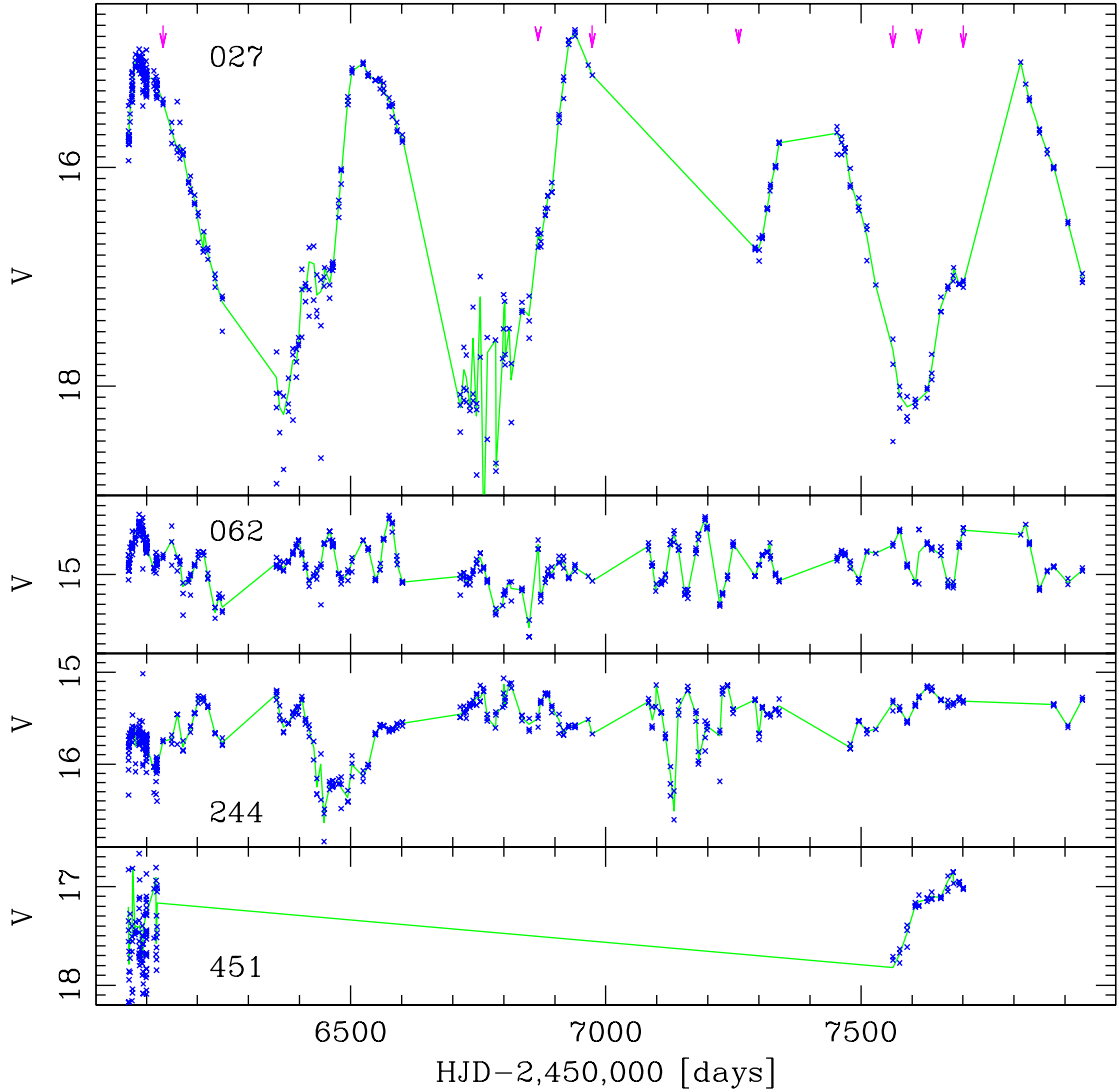}
\caption{The $V$-band light curves of stars \#27, \#62, \#244, and \#451 (BGSU + P1AU only) are shown using the symbols from the previous figures.
  \label{fig_lc4vars}}
\end{figure}

In Figure~\ref{fig_lc4vars}, we show the $V$ magnitude as a function of time for several variable stars that display a range of properties. 
 Most of the variables discussed in \citet{wehrung13} are labeled ``[WLR2013]~\#'' in the AAVSO's Variable Star Index (VSX) using the same identification number  listed herein.

A table showing the median and extreme light curve values in each bandpass is presented in the Supplement. The extrema were computed at the 2\% and 98\% points of the magnitude-sorted data sets to reduce the effects of outlier points and photometric scatter. We define the range of variation, $R_f$, as the difference between the extrema in a particular filter $f$. In many studies this is termed the amplitude, but since many of the stars studied here have irregular behaviors, we reserve that term for the sinusoids in the Fourier period analysis. As is usually seen for pulsating stars, we found that $R_V > R_I$ for most of the stars in our sample.

Upon inspecting the light curves, it is clear that most of the stars are red, long period variables similar to those observed in the Large Magellanic Cloud by \citet{soszynski13}. We present in Table~\ref{tab_per} the three periods ($P_1$, $P_2$, and $P_3$ in days) and their sinusoidal half-amplitudes ($A_1$, $A_2$, and $A_3$ in mag) with the largest power as determined from the $V$ data for each star using the discrete Fourier transform software \textsc{VStar} \citep{benn12}. We excluded from Table~\ref{tab_per} periods that are aliases of stronger periods. The semi-amplitudes are closely correlated with the power metric output by \textsc{VStar}, so larger amplitudes indicate higher power and more likely periods. We include the range, $R_V$, in the second column of Table~\ref{tab_per} for comparison with the amplitudes. In some cases, a light curve modeled using the three highest-power sinusoids does not account for all the variation seen in the observed light curve, indicating additional sources of natural variation that may include more periods operating in the star, the presence of truly irregular (chaotic) pulsation, magnetic starspot activity, convective cells, or absorption by varying amounts of circumstellar dust. Other stars, like \#27, have a single period that dominates the power spectrum. 
The periods listed in bold are the ones we find are most successful in characterizing each star's light curve. 

In the following subsections, we highlight several stars in each of four broad categories. A detailed discussion of each star, along with $V$-band light curves of stars not shown in Figure~\ref{fig_lc4vars}, are presented in the Supplement.

\begin{deluxetable}{lrrrrrrrcl}
\tablecaption{Results of Fourier Period Analysis \label{tab_per}}
\tablehead{
\colhead{Star} & \colhead{$R_V$} & \colhead{$P_1$} & \colhead{$A_1$} & 
\colhead{$P_2$} & \colhead{$A_2$} & \colhead{$P_3$} & \colhead{$A_3$} & 
\colhead{Type} & \colhead{Comments}   
}
\startdata
417 & 1.03 & 423 & 0.459 & 327 & 0.174 & 602 & 0.134 & NC & V1835~Aql,  $P_{\rm orb} = 419$~d\\
001 & 0.45 & \textbf{49.5} & 0.061 & \textbf{34.8} & 0.049 & 47.3 & 0.043 & SRB & ASAS, 334-d LSP? \\
026 & 0.58 & \textbf{70.2} & 0.110 & \textbf{73.6} & 0.088 & \textit{348} & 0.104 & SRB & LSP \\
027 & 3.31 & \textbf{426} & 1.599 & 586 & 0.900 & 281 & 0.718 & M & ASAS-SN  \\
062 & 0.91 & \textbf{61.8} & 0.193 & \textit{514} & 0.144 & 283 & 0.109 & SRA & LSP  \\
085 & 0.40 & 42 & 0.090 & \nodata & \nodata & \nodata & \nodata & SR & P1AU  \\
132 & 0.42 & 43 & 0.117 & \nodata & \nodata & \nodata & \nodata & SR & P1AU  \\
244 & 1.18 & \textit{471} & 0.244 & \textit{337} & 0.242 & \textbf{67.3} & 0.128  & SRB? & 2 LSP?  \\
255 & 0.48 & \textit{603} & 0.066 & \textbf{45.4} & 0.059 & \textbf{35.4} & 0.045 & SRB & LSP  \\
419 & 1.83 & 185 & 0.692 & \nodata & \nodata & \nodata & \nodata & SRA & P1AU  \\
991 & 0.28 & 78.6 & 0.064 & 44.2 & 0.026 & 46.4 & 0.023 & ELL & $P_{\rm orb} = 157.2$~d  \\
992 & 0.71 & \textbf{38.1} & 0.086 & 104.4 & 0.085 & \textit{685} & 0.091 & SRB & Also \textbf{ 51.3~d}  \\
993 & 0.20 & 0.93551 & 0.047 & 15.109 & 0.043 & 0.48270 & 0.037 & ? & -  \\
995 & 0.15 & \textit{602} & 0.016 & \textbf{21.1} & 0.013 & \textit{282} & 0.013 & SRB? & 2 LSP?  \\
996 & 0.47 & \textbf{42.0} & 0.080 & \textit{310} & 0.067 & \textbf{42.8} & 0.065 & SRB & LSP, ASAS-SN  \\
987 & 0.72 & 0.18296 & 0.170 & 1.6093 & 0.092 & 8.081 & 0.065 & EW & $P_{\rm orb} = 0.36592$~d  \\
988 & 0.49 & \textbf{41.8} & 0.072 & \textbf{32.9} & 0.0.071 & 104 & 0.063 & SRB & 699-d LSP?  \\
105 & 0.27 & 24 & 0.046 & \textit{139} & 0.054 & \nodata & \nodata & SR & P1AU, LSP?, ASAS-SN  \\
154 & 0.48 & 43 & 0.105 & 73 & 0.101 & \nodata & \nodata & SRB & P1AU  \\
184 & 0.36 & 54 & 0.102 & \nodata & \nodata & \nodata & \nodata & SR & P1AU  \\
270 & 1.97 & 79.6 & 0.659 & \nodata & \nodata & \nodata & \nodata & SRA & P1AU, ASAS-SN  \\
358 & 0.52 & \textit{243} & 0.204 & 60 & 0.181 & \nodata & \nodata & SR & P1AU, LSP?  \\
451 & 1.15 & $>$250 & $>$0.48 & \nodata & \nodata & \nodata & \nodata & SRA & P1AU  \\
 \enddata
\tablecomments{Periods in bold-face indicate most likely pulsation periods, while periods in italics are likely to be long secondary periods (LSPs).}
\end{deluxetable}

\subsection{High amplitude, regular LPVs}

Variable star \#27 has an extremely red mean color  and was detected by the Infrared Astronomical Satellite (IRAS). We found it to have a large $V$-band range of 3.31~mag and slow variations as shown in Figure~\ref{fig_lc4vars} with a \textsc{VStar} period of 426~d, indicating that it is a Mira-type LPV.
This star was also detected by the All-Sky Automated Survey for Supernovae \citep{shappee14} as ASASSN-V J190737.43+000609.1 and \citet{jayasinghe18} classified it as a semi-regular variable with a period of 414.94~d,  mean $V = 15.42$~mag and amplitude $\Delta V = 1.05$~mag.  We suspect that the low spatial resolution of their survey resulted in blending of light with nearby stars, causing a magnitude 0.7~mag brighter than ours and a diluted amplitude. We assert that our photometric properties and classification as a Mira variable are preferrable. 

Star \#62 shows a relatively large range of $R_V = 0.91$~mag and regular pulsation in Figure~\ref{fig_lc4vars}. We find
two dominant periods at 61.8 and 514 days, which may represent the primary pulsation period and a long secondary period (LSP), respectively. We classify this star as a classical semi-regular LPV (SRA).

\subsection{Lower amplitude, multi-periodic LPVs}

Many of the variable stars have red colors but lower photometric ranges ($R_V$) and show irregular light curve behavior in which \textsc{VStar} finds evidence for multiple periods with low-amplitudes. We classify these stars (\#1, \#26, \#244, \#255, \#992, \#995, \#996, and \#988) as semi-regular LPVs of the SRB type. Many of them show evidence of long secondary periods, as noted in the comments column of Table~\ref{tab_per}.  

Several of these stars are noteworthy. For example, the bright variable \#1 shows evidence for three periods in the 30-50~d range, one of which (34.7~d) is confirmed using data from the All-Sky Automated Survey \citep{pojmanski2004}. A lower-power period of $\sim$334~d also appears in both data sets, which may be an LSP.  

Two stars, \#244 (see Figure~\ref{fig_lc4vars}) and \#995, have two relatively strong periods in the 300-600~d range, where one more typically finds higher amplitude, Mira-like pulsation
\citep{soszynski13}. We speculate that these stars may exhibit two LSPs each. While we are not aware of any such stars in the literature, it is a possibility under the hypothesis promoted by \citet{sosz21} that the LSP phenomenon is due to obscuration by a dust cloud surrounding and trailing a very low-mass companion orbiting the red giant. In the cases of stars \#244 and \#995, we propose the presence of two such companions with orbital periods corresponding to the LSPs of each star.

Variable \#996 appears to be a multi-periodic SRB with an LSP. It was also detected by \citet{jayasinghe18} as  ASASSN-V J190742.80+000440.5 and classified as an SR-type LPV with a period of 29.23~d, mean $V = 15.19$~mag, and amplitude $\Delta V = 0.28$~mag. Our \textsc{VStar} analysis found a period of 29.3~d at the eighth-ranked peak in our power spectrum, raising the possibility that pulsation activity shifts energy between different modes over time in this star.  As was the case for star \#27, our median magnitude ($V$ = 15.44~mag) is fainter and our range is larger, consistent with blended light affecting the ASAS-SN photometry of \#996.

\subsection{Bluer, non-LPV variables}

Three variables shown in the CMD (Figure~\ref{fig_cmdvars}) are significantly bluer than the other variables found in this field. Star \#991, the reddest of the three, appears to be a ellipsoidal variable (ELL) with an orbital period of 157~d and $E_0 = 2,456,803$~d, composed of at least one red giant.   Star \#987 is a contact eclipsing binary (W~Ursae Majoris, EW-type) with an orbital period of 0.36592~d and $E_0 = 2,456,803.425$~d. Variable \#993 is extremely blue, but classifying the star definitively proved difficult: the phased light curves are equally good for a pulsating variable with a period of 15.11~d ($E_0 = 2,456,803.7$~d) or of 0.93551~d, or an ellipsoidal/binary star with an orbital period of 1.87102~d ($E_0 = 2,456,802.94$~d). More data taken with a higher sampling rate are needed to clarify the nature of \#993, and spectra could indicate whether it is a S~Doradus star, Wolf-Rayet star, or luminous blue variable. Light curves phased with theses ephemerides are shown in the Supplement.

\subsection{Variables in the wider fields} 

Though the lower spatial resolution and shorter time span of the images in the BGSU and P1AU data sets make them less-suitable for detecting variables than the P5AA set, their fields of view cover more than five times the area so their analysis yielded additional variable stars. The bottom panel of Figure~\ref{fig_lc4vars}  shows the relatively short time span of each of these data sets and the long gap between them. %
Given the reduction in data, we focused our detection on stars with slow, correlated variations in $V$ and $I$, and our \textsc{VStar}  analysis on the strongest period in each star. We were also less ambitious in our effort to characterize each star's variability type, and acknowledge that many variables with shorter periods remain undetected in this region.

In the original \citet{wehrung13} paper, we detected three LPVs in this region, \#85, \#132, and \#419, but were unable to characterize their behaviors. With the additional data from the P1AU images, we were able to obtain period estimates for all three. The former two stars have short periods of 42-43~d and can be classified as semi-regular (SR) variables. Data for \#419 suggests a period of $\sim$185~d and a large amplitude near the $R_V = 2.5$~mag border that separates SRA from Mira LPVs, though for now we classify it as SRA type.

Among the six new variables detected in the P1AU data set, four appear to be SR or SRB types with modest photometric ranges and evidence of one or more periods (\#105, \#154, \#184, and \#358). Two stars (\#270 and \#451) have longer periods and larger ranges, near the borderline between SRA and Mira LPVs, but we classify them provisionally as SRA until photometry across more pulsation cycles is obtained. Our light curve for star \#451 is shown in Figure~\ref{fig_lc4vars}.

Two of these six variables were also detected by the All-Sky Automated Survey for Supernovae \citep{shappee14}. Our star \#270 is the same as ASASSN-V J190726.81-000347.7, which \citet{jayasinghe18} found to have mean $V = 15.09$ mag, amplitude $\Delta V = 1.15$ mag, period 79.51 d and to be a semi-regular variable. Our periods agree extremely well, but as in previous cases, the ASAS-SN mean magnitude is brighter and amplitude is smaller than ours, suggesting that stellar image blending has affected their photometry.
Star \#105 is ASASSN-V J190813.72+000019.4 and \citet{jayasinghe18} list it with mean $V$ = 14.77~mag, $V$ amplitude 0.27~mag and an irregular ``L'' classification with no discernible period. The ASAS-SN photometry agrees well with ours for this star, which has no bright stars nearby on our images that might effect the photometry. Our data shows low-amplitude periodic behavior at 24~d and 139~d, enabling us to classify it as an SR pulsator.

\section{Summary}    \label{sec_concl}

We obtained time-series CCD photometry in the $BVRI$ passbands of the recently identified symbiotic nova V1835~Aql over an interval of 5.1~years. The star varies through a magnitude range of $R_V \approx 0.9$~mag and $R_I \approx 0.3$~mag with a regular cycle described by the ephemeris $JD_{min} = (419 \pm 10)~N_{cyc} + (2,456,834 \pm 10$)~d. This orbital period is relatively short for a SySt, ranking eighth of 53 objects in the distribution given by \citet{grom13}. No sharp-edged stellar eclipses are evident. We see a dip at $\phi = 0.5$ in the light curve shown in Figure~\ref{fig_lc_phi2} that we attribute to tidal distortion of the giant star and the consequent ellipsoidal effect on the light curve. This hints that V1835~Aql may transfer some of its mass to the hot component via Roche lobe overflow \citep{miko2003b}, rather than the more widely-accepted model of stellar wind transfer. We show that any intrinsic variation of the red giant, for example caused by pulsation, is weak; this, and the dereddened near-infrared color of the red giant, are consistent with V1835~Aql being a S-type (stellar) SySt. The strong photometric modulation of the object in the bluer passbands suggests that the compact stellar object that irradiates the giant star's wind has a relatively high luminosity, consistent with fusion of accreted hydrogen continually as it reaches the stellar surface: a \textit{burn}-SySt. It is noteworthy that V1835~Aql is among a small number of SySt that have been observed to undergo a major nove-like outburst \citep{williams2005}.

We also provide $BVRI$ time series photometry on 22 variable stars in the field around V1835~Aql, nine of which were initially detected in \citet{wehrung13}. We list the stars' photometric properties and periods found using Fourier analysis. Most turn out to be long period variable red giants, including one Mira, four semi-regular (SRA), nine multi-periodic (SRB), and five LPVs which showed convincing periodic behavior for which we did not have enough data to classify with confidence (SR). Many of these stars have long secondary periods (LSPs) and two stars appear to have two LSPs each, which seems viable under the hypothesis that the LSP behavior is caused by dimming due to dust clouds entrained by orbiting, low-mass companion objects \citep{sosz21}. Detailed analysis of each star and light curves are provided in the Supplement.

We recommend that V1835~Aql receive additional photometric observations in the near ultraviolet and near infrared regions (e.g., $UB,JHK$) in order to confirm the ellipsoidal variations suspected herein, and to quantify its physical properties. Spectroscopic radial velocity measurements over all phases of the light curve would also confirm the orbital period and perhaps provide a mass ratio as shown in \citet{miko2003a} %
We also note that many of these stars, including V1835~Aql, are receiving ongoing optical photometric monitoring by the Zwicky Transient Facility \citep{ztf19} and the Gaia astrometric satellite \citep{GaiaSat}. 

\acknowledgments
Acknowledgements: 
This work represents partial fulfillment of a Master of Science degree in Physics at BGSU for RVC.   We thank BGSU undergraduate Bianca Legeza for help in obtaining and analyzing data from online sources.
We thank Howard Bond and an anonymous reviewer for helpful feedback on our manuscript.
This research was made possible through the use of the AAVSO Photometric All-Sky Survey
(APASS), funded by the Robert Martin Ayers Sciences Fund, as well as the AAVSO's  Variable Star Index (VSX). We thank the Robert Martin Ayers Science Fund for sponsoring our observing time on the PROMPT telescopes for this project. 
This work has made use of data from the European Space Agency (ESA) mission
{\it Gaia} (\url{https://www.cosmos.esa.int/gaia}), processed by the {\it Gaia}
Data Processing and Analysis Consortium (DPAC,
\url{https://www.cosmos.esa.int/web/gaia/dpac/consortium}). Funding for the DPAC
has been provided by national institutions, in particular the institutions
participating in the {\it Gaia} Multilateral Agreement. 
This research utilized the Image Reduction and Analysis Facility (IRAF), which was developed and distributed by NOAO through AURA, Inc. under agreement with the NSF.

\vspace{5mm}
\facilities{AAVSO, CTIO:PROMPT}

\software{IRAF (\url{https://iraf.net/})
\software{\textsc{VStar} (https://www.aavso.org/vstar) }
          }

\bibliographystyle{ieee}

\end{document}


\title{Optical Time-Series Photometry of the Symbiotic Nova V1835~Aquilae\\
Supplemental Data: details of variable stars}

\correspondingauthor{Andrew Layden}
\email{laydena@bgsu.edu}

\author[0000-0002-4475-3181]{Robert V. Caddy}
\affil{University of Pittsburgh\\
100 Allen Hall, Department of Physics \& Astronomy \\
Pittsburgh, PA 15261, USA}
\altaffiliation{Bowling Green State University\\
104 Overman Hall, Physics \& Astronomy Department\\
Bowling Green, OH 43403, USA}

\author[0000-0002-6345-5171]{ Andrew C. Layden}
\affiliation{Bowling Green State University\\
104 Overman Hall, Physics \& Astronomy Department\\
Bowling Green, OH 43403, USA}

\author{Bianca S.~N. Legeza}
\affiliation{Bowling Green State University\\
104 Overman Hall, Physics \& Astronomy Department\\
Bowling Green, OH 43403, USA}

\begin{abstract}

We present detailed descriptions of each of 22 variable stars, including the \textsc{VStar} period analysis, light curve characteristics, and mean color, to help determine each star's variability type. The symbiotic nova V1835~Aql was discussed in the main text, so is not discussed herein.

\end{abstract}

\section{Overview}  \label{sec_intro}

For each of the 22 variable stars discussed in Section~5 of the main paper, we determined 
the median and extreme light curve values in each bandpass, shown here in Table~\ref{tab_lcparams}. The extrema were computed at the 2\% and 98\% points of the magnitude-sorted data sets to reduce the effects of outlier points and photometric scatter. The range of variation in filter $f$, $R_f$, 
is the difference between the extrema; we reserve the term amplitude for the sinusoids in the Fourier period analysis. 

As described in the main text, we used the discrete Fourier transform method implemented in \textsc{VStar} to identify periods in our time-series data. We began the analysis of each star by plotting the $V$ and $I$ magnitudes as a function of time (the data are available electronically through Table~3 of the main paper). Stars with slow light variation like long period variables (LPVs) are apparent in such plots, and so we proceeded to run a \textsc{VStar} search on the interval 10-1000 days with 0.1-day resolution on the $V$-band data, and focused attention on the periods with the highest power. We identified aliases in the usual way, and removed the period in an alias-pair that had the lower power from further consideration. We selected the three periods with the highest power as most likely including the actual period of variation of the star. For stars whose time-magnitude plots showed scatter, consistent with a period shorter than our sampling rate, we searched for periods using the interval described above, then tested a 1-10~d interval with 0.0001~d resolution, and then 0.1-1.0~d at 0.00001~d.

In the first of the following sections, we consider the stars identified in the first paper in this series; their light curves are shown in Figure~\ref{fig_lc1}. In Section~3 we discuss the variables that were discovered in the new PROMPT~4 data sets, with their light curves are shown in Figures~\ref{fig_lc1} and \ref{fig_phi}. Finally, we cover the most convincing variable star candidates found in the PROMPT~1 data set in Sec.~4, with light curves in Figure~\ref{fig_lc3}.

\begin{longrotatetable}
\begin{deluxetable}{lrrrrrrrrrrrrrrrrrrr}
\tablecaption{Photometric Properties of the Variable Stars \label{tab_lcparams}}
\tablehead{
\colhead{Star} & 
\colhead{$V_{\rm{med}}$} & \colhead{$V_{\rm{max}}$} & \colhead{$V_{\rm{min}}$} & \colhead{$N_V$} & \colhead{$I_{\rm{med}}$} & \colhead{$I_{\rm{max}}$}  & \colhead{$I_{\rm{min}}$} & \colhead{$N_I$} & 
\colhead{$R_{\rm{med}}$} & \colhead{$R_{\rm{max}}$}  & \colhead{$R_{\rm{min}}$} & \colhead{$N_R$} & 
\colhead{$B_{\rm{med}}$} & \colhead{$B_{\rm{max}}$}  & \colhead{$B_{\rm{min}}$} & \colhead{$N_B$}  
}
\startdata
417 & 16.007 & 15.55 & 16.58 & 335 &  12.842 & 12.72 & 13.03 & 332 & 14.257 & 14.10 & 14.99 &  99 & 18.027 & 17.65 & 18.79 & 78 \\
001 & 11.728 & 11.50 & 11.94 & 320 &   7.390 &  7.26 &  7.54 & 277 &   9.813 &  9.66 &  9.90 &  97 & 14.300 & 14.13 & 14.53 & 82 \\
026 & 13.864 & 13.54 & 14.12 & 329 &  10.118 &  9.97 & 10.22 & 314 &  12.153 & 11.88 & 12.47 &  92 & 16.414 & 16.06 & 16.71 & 82 \\
027 & 16.152 & 14.80 & 18.11 & 198 &  10.175 &  8.85 & 11.35 & 316 & 13.068 & 12.13 & 14.0~ &  97 & \nodata &\nodata & \nodata & 0 \\
062 & 14.917 & 14.47 & 15.38 & 352 &  10.791 & 10.62 & 10.98 & 329 &  13.020 & 12.68 & 13.32 &  99 & 17.469 & 17.06 & 17.78 & 82 \\
085 & 14.960 & 14.79 & 15.19 & 149 &  10.863 & 10.75 & 10.96 & 244 & \nodata & \nodata & \nodata & 0 & \nodata & \nodata & \nodata & 0 \\
132 & 15.524 & 15.37 & 15.79 & 115 &  11.020 & 10.98 & 11.15 & 214 & \nodata & \nodata & \nodata & 0 & \nodata & \nodata & \nodata & 0 \\
244 & 15.469 & 15.15 & 16.32 & 314 &  11.839 & 11.67 & 12.12 & 302 &  13.954 & 13.66 & 14.54 &  93 & 18.150 & 17.60 & 18.94 & 74 \\
255 & 15.769 & 15.55 & 16.03 & 335 &  12.065 & 11.98 & 12.18 & 322 &  14.088 & 13.85 & 14.20 &  97 & 18.236 & 18.00 & 18.68 & 75 \\
419 & 16.667 & 16.18 & 18.01 &  97 &  12.380 & 12.33 & 13.36 & 244 & \nodata & \nodata & \nodata & 0 & \nodata & \nodata & \nodata & 0 \\
991 & 13.912 & 13.80 & 14.07 & 334 &  12.013 & 11.96 & 12.12 & 315 &  12.950 & 12.86 & 13.06 &  97 & 15.668 & 15.56 & 15.84 & 82 \\
992 & 16.781 & 16.36 & 17.06 & 278 &  12.226 & 12.08 & 12.42 & 332 &  14.709 & 14.46 & 14.85 &  99 & \nodata & \nodata & \nodata & 0 \\
993 & 13.096 & 13.01 & 13.21 & 319 &  12.294 & 12.23 & 12.39 & 303 &  12.709 & 12.63 & 12.78 &  92 & 13.821 & 13.72 & 13.90 & 82 \\
995 & 11.933 & 11.85 & 12.00 & 341 &   9.121 &  9.04 &  9.16 & 192 &  10.615 & 10.54 & 10.64 &  93 & 14.186 & 14.10 & 14.24 & 82  \\
996 & 15.442 & 15.18 & 15.64 & 348 &  11.525 & 11.44 & 11.61 & 329 &  13.695 & 13.52 & 13.85 &  99 & 18.042 & 17.84 & 18.21 & 80 \\
987 & 16.211 & 15.85 & 16.58 & 198 &  14.856 & 14.64 & 15.18 & 224 &  15.533 & 15.25 & 15.88 &  89 & 17.338 & 17.15 & 17.67 & 81 \\
988 & 16.063 & 15.80 & 16.29 & 226 &  12.365 & 12.27 & 12.47 & 255 &  14.351 & 14.19 & 14.58 &  89 & 18.492 & 18.22 & 18.77 & 68 \\
105 & 14.628 & 14.51 & 14.78 & 150 &  11.196 & 11.13 & 11.26 & 244 & \nodata & \nodata & \nodata & 0 & \nodata & \nodata & \nodata & 0 \\
154 & 15.215 & 14.98 & 15.46 & 148 &  11.010 & 10.95 & 11.23 & 245 & \nodata & \nodata & \nodata & 0 & \nodata & \nodata & \nodata & 0 \\
184 & 15.085 & 14.91 & 15.27 & 122 &  11.663 & 11.56 & 11.72 & 218 & \nodata & \nodata & \nodata & 0 & \nodata & \nodata & \nodata & 0 \\
270 & 15.179 & 14.38 & 16.35 & 145 &  11.813 & 11.69 & 12.44 & 244 & \nodata & \nodata & \nodata & 0 & \nodata & \nodata & \nodata & 0 \\
358 & 16.488 & 16.20 & 16.72 &  52 &  11.801 & 11.57 & 11.87 &  78 & \nodata & \nodata & \nodata & 0 & \nodata & \nodata & \nodata & 0 \\
451 & 17.116 & 16.85 & 18.0  &  34 &  11.630 & 11.45 & 12.0  & 245 & \nodata & \nodata & \nodata & 0 & \nodata & \nodata & \nodata & 0 \\
\enddata
\end{deluxetable}
\end{longrotatetable}

\section{Variables reported by Wehrung~et~al.~(2013) }  \label{sec_wlr13}

\begin{figure}
\plotone{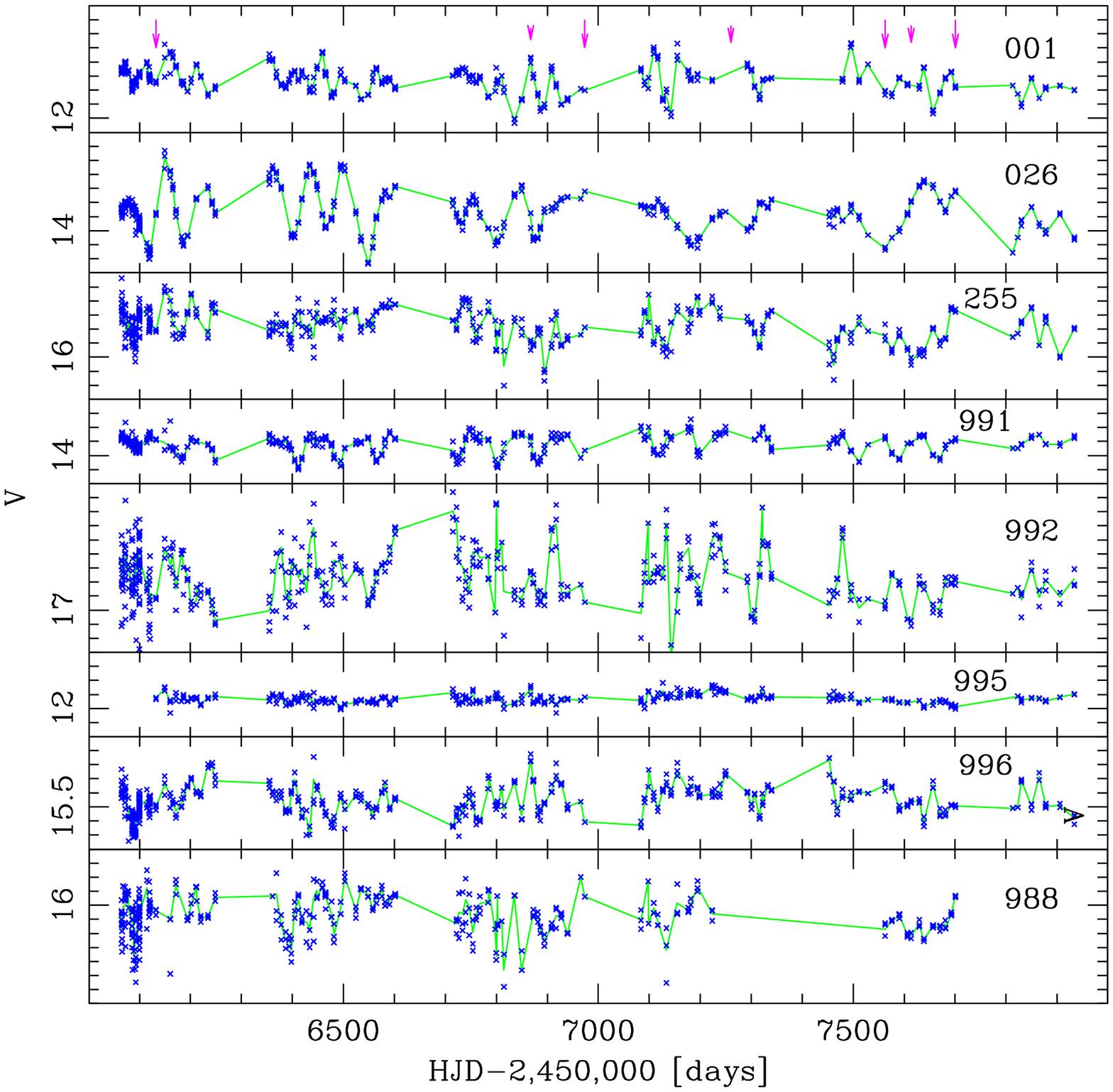}
\caption{The  $V$-band light curves of variable stars, listed by their identification number. Blue crosses are from individual images, and green lines connect nightly mean points. Small tick marks are 0.1~mag on the vertical scale. 
  \label{fig_lc1}}
\end{figure}

\begin{figure}
\plotone{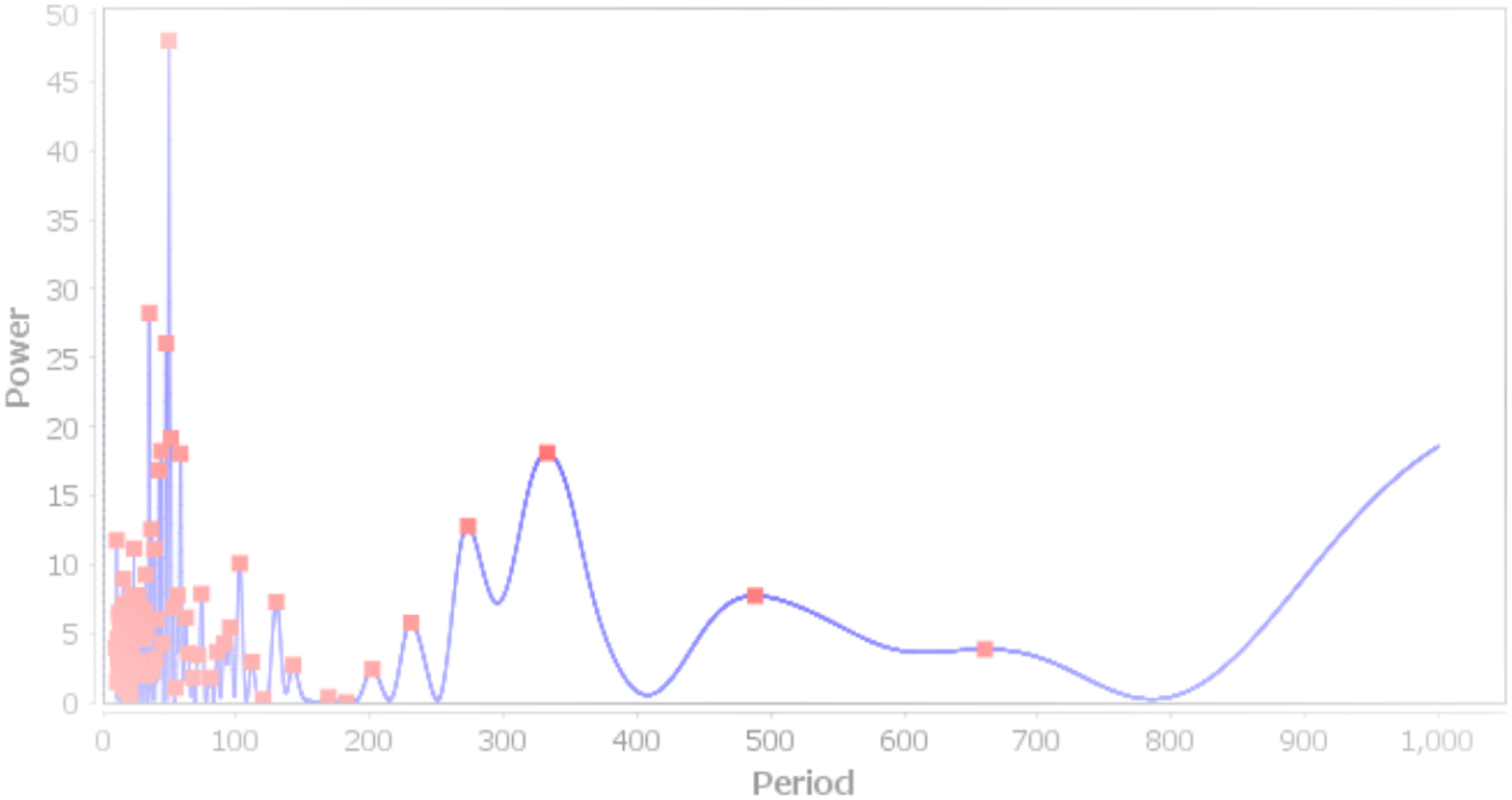}
\caption{The power spectrum from \textsc{VStar} using our $V$-band data for star \#001 is shown as an example of our period detection methodology for long-period variable stars. 
  \label{fig_power}}
\end{figure}

\textbf{Star \#1}, listed in the VSX as ``[WLR2013]~1,'' has an irregular appearing light curve with a range of 0.45 mag in $V$ and a rather red mean color of $\langle V \rangle - \langle I \rangle = 4.34$~mag (see Figure~\ref{fig_lc1}). The top three periods found using \textsc{VStar} on our $V$-band data (49.5, 34.8, and 47.3 days) all have significant power (the power spectrum from \textsc{VStar} is shown in Figure~\ref{fig_power}), but  when modeled do not account for all the variation seen in the observed light curve, suggesting that more periods or causes of variability are at play. This star was also observed by the All-Sky Automated Survey (ASAS, \citet{pojmanski2004}). A \textsc{VStar} analysis of that data yielded 289.8~d, 31.6~d, and 34.6~d, all with semi-amplitudes of $\sim$0.055~mag. The 34-day period appears in both data sets, and the 31-d period is an alias of this value. A longer, 334-day period shows up with an amplitude of $\sim$0.04~mag in both data sets, and may represent a long secondary period (LSP). Star~\#1 appears to be a multi-periodic semi-regular LPV (SRB), with periods of 49.5 and 34.7 days expressed most strongly. We calculate the absolute magnitude of star~\#1 to be $M_{\rm V} = -0.1$~mag based on the median magnitude in Table~1 and the distance in Table~2 of the main paper, along with $E(B-V) = 0.67$~mag.

\textbf{Star \#26:} This variable star, called ``[WLR2013]~26'' in the VSX, is slightly bluer than Star~\#1 at $\langle V \rangle - \langle I \rangle = 3.75$ mag, and has a larger range of 0.58~mag in $V$. The three non-aliased periods with the highest power are 70.2, 73.6 and 348 days. The latter is likely to be a LSP, and the 70.2 d period is the dominant period, describing much of the observed variation, making star \#26 a rather classical semi-regular variable. When added to the \textsc{VStar} model, the 73.6~d period interacts with the 70.2~d period to account for some the amplitude variation seen in the light curve, supporting the idea that this star is truly multi-periodic, and so we classify it as SRB. We estimate $M_{\rm V} = -1.2$~mag for star~\#26 using the data described in the previous paragraph.

\textbf{Star \#27:} This variable did not appear in the VSX listing with the other WLR13 stars. It is extremely red, $\langle V \rangle - \langle I \rangle = 5.98$ mag, and was detected by the IRAS infrared satellite. We found it to have large $V$-band range of 3.31~mag and slow variations with a \textsc{VStar} period of 426~d, as shown in Figure~5 of the main text, properties that indicate it is a Mira-type LPV.
Star \#27 was also detected by the All-Sky Automated Survey for Supernovae \citep{shappee14} which observed in the $V$-band with a 2-3~d cadence between Julian dates 2,457,100 and 2,458,400 days, as ASASSN-V J190737.43+000609.1. \citet{jayasinghe18} classified the star as a semi-regular variable with a period of 414.94~d, a mean $V = 15.42$~mag and amplitude $\Delta V = 1.05$~mag from observations spanning over two full cycles.  While our periods agree well, we suspect that the low spatial resolution of their survey resulted in blending of light with nearby stars, causing a brighter magnitude (we found a median $V$ of 16.15~mag) and diluted amplitude, and we therefore assert that our photometric properties and classification as a Mira variable are correct. We estimate $M_{\rm V} = +1.6$~mag for star~\#27, which seems faint for an LPV; we look to improved parallaxes in upcoming \textit{Gaia} data releases to resolve this issue.

\textbf{Star \#62:} This variable, named ``[WLR2013]~62'' in the VSX, presents two strong periods in \textsc{VStar} at 61.8 and 514 days, which account for most of the light curve variations and likely represent the primary pulsation period and an LSP. This star is rather red, $\langle V \rangle - \langle I \rangle = 4.13$ mag, and has a relatively large range in $V$ of 0.91 mag, characteristic of a classical semi-regular LPV (SRA). The star's strongly periodic nature, modulated perhaps by other periods (64.8 and 55.3~d have moderate power in our analysis), is evident in the light curve shown in Figure~5 of the main text. We estimate $M_{\rm V} = -0.9$~mag for star~\#62.

\textbf{Star \#244:} The color and range of this variable, also known as ``WLR2013]~244,'' are 3.63 and 1.18 mag respectively, similar to that of \#62. The light curve of star \#244, which is also shown in Figure~5, has a deep and prolonged dip at JD $\approx$ 2,456,500~d that could be attributed to a dust obscuration episode. The \textsc{VStar} analysis indicates strong periods at 471 and 337 days whose minima coincide at this dip. Ignoring two additional periods at 179 and 210~d which are close to aliases of the first two periods, the next strongest period is at 67.3~d. Including this in a model with 471 and 377~d accounts for much of the variability seen in the light curve, though we note that a number of other periods are present with similar power, including one at 117~d that improves the modeled light curve. We propose that the 67.3~d period is the dominant pulsation period of this semi-regular LPV, and that two LSPs are present in this star. While we are not aware of any LPVs with multiple LSPs, it is a possibility under the hypothesis promoted by \citet{sosz21} that the LSP phenomenon is due to obscuration by a dust cloud surrounding and trailing a very low-mass companion orbiting the red giant. In the case of star \#244, we propose the presence of two companions with orbital periods of 377 and 471~d, respectively.

\textbf{Star \#255} or ``[WLR2013]~255,'' is similar to \#244 in color (3.70 mag), but has a smaller $R_{\rm V} = 0.48$~mag (see Figure~\ref{fig_lc1}). The two strongest periods of 603 and 45.4~d likely indicate the LSP and pulsation periods of this semi-regular variable. The 35.4~d period and several others of similar length and power may modulate the amplitude over time, and so we classify this star as an SRB.

The \citet{wehrung13} stars \#85, 132 and 419 had relatively few observations because they were outside the narrower PROMPT fields of view, and were only observed using the BGSU and P1AU telescopes. They are discussed in Sec.~\ref{sec_wide}.

\section{New variables detected by the narrow field imagers} \label{sec_narrow}

\begin{figure}
\plotone{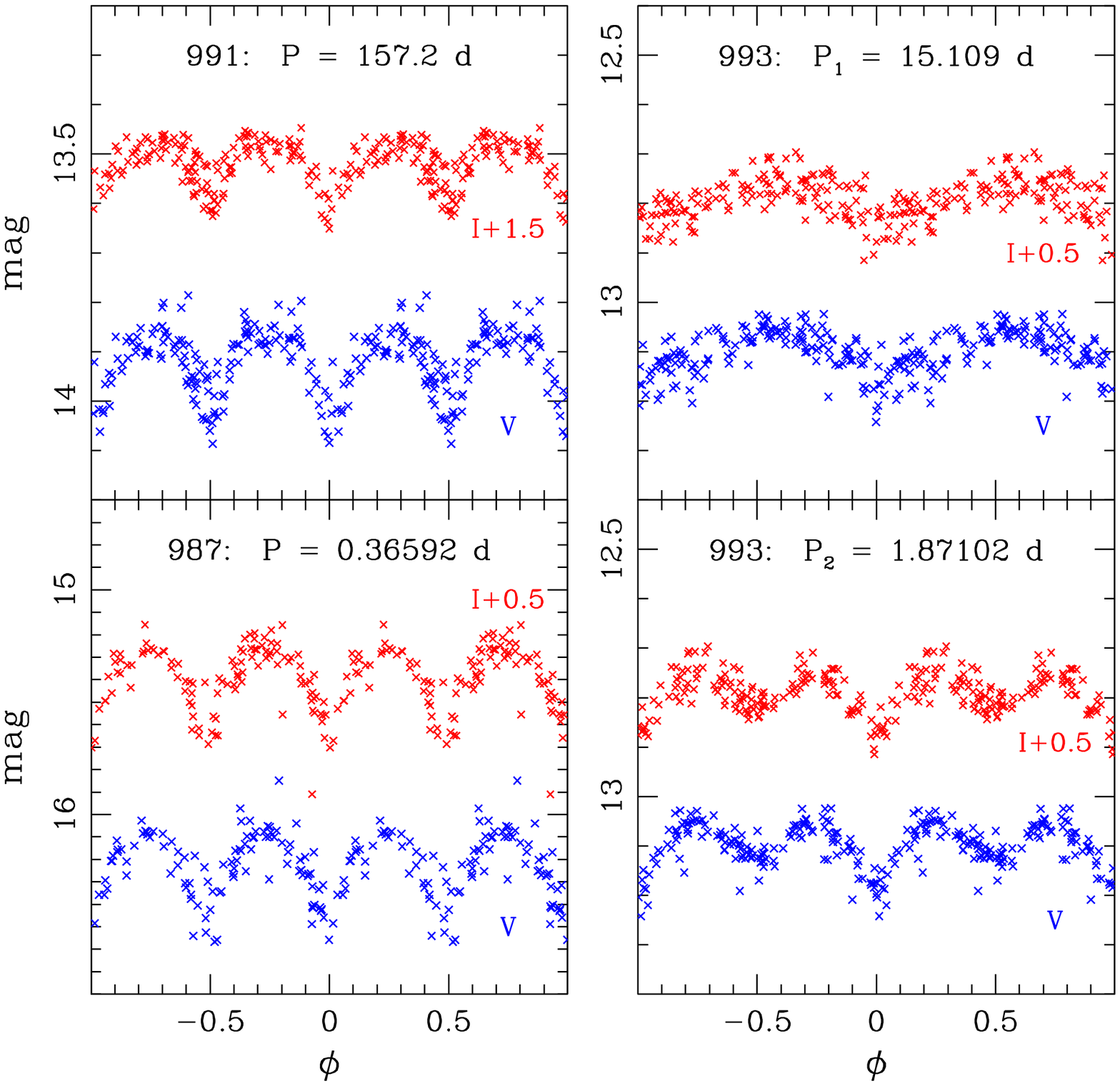}
\caption{The  $V$ and $I$ light curves of variable stars \#991, \#993, and \#987 phased  using the periods indicated. Crosses mark nightly average magnitudes; $I$ data are shifted by the amounts shown for convenience of display.
  \label{fig_phi}}
\end{figure}

\begin{figure}
\plotone{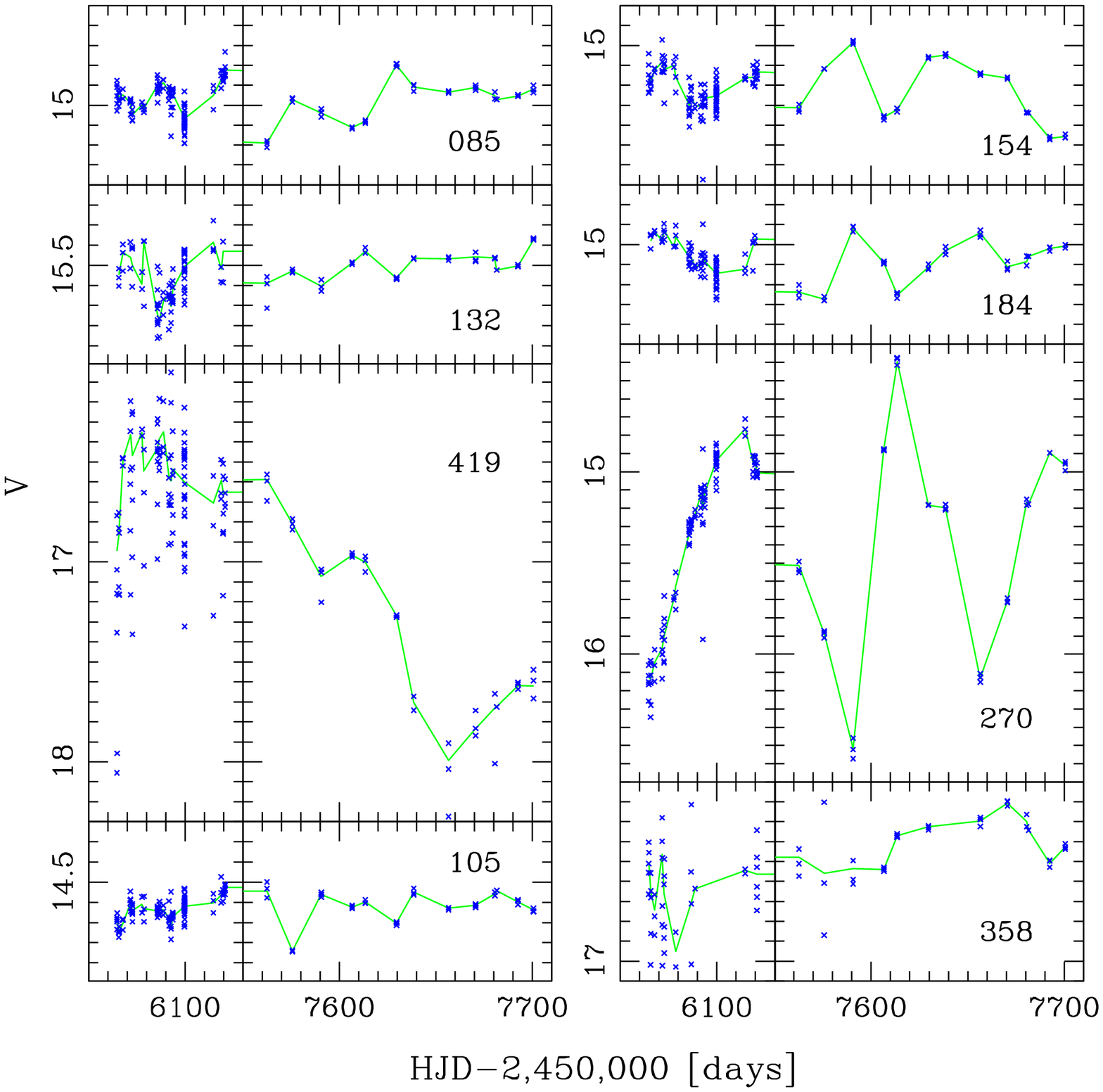}
\caption{The  $V$-band light curves of variable stars found in the wide-field images, listed by their identification number. Symbols are as in Figure~\ref{fig_lc1}. Small tick marks are 0.1~mag on the vertical scale. For each star, the left panel shows BGSU data, and the right panel shows data from the Prompt~1 telescope.
  \label{fig_lc3}}
\end{figure}

\textbf{Star \#991:} At  $\langle V \rangle - \langle I \rangle = 1.90$ mag, this newly-detected variable star is significantly bluer than any of the stars described in the previous section, appearing near the blue edge of the non-variable red giant stars shown in Figure~1 of the main text. 
Its $V$-band light curve is quite regular, and our \textsc{VStar} analysis shows that a single period at 78.6~d describes most of the observed variation. The light curve phased with twice this period, to represent an orbital period, looks similar to the light curves of ellipsoidal red giants detected in the Large Magellanic Cloud shown in Figure~1 of \citet{soszynski04}. The period, $I$-band amplitude, and dereddened color are also consistent with the distributions of these properties in the sample of \citet{soszynski04}. We therefore classify star \#991 as an ellipsoidal (ELL-type) red giant variable. We estimate $M_{\rm V} = -0.4$~mag, and show its phased light curve in Figure~\ref{fig_phi} using the  ephemeris $JD_{min} = (157.2) \times N_{cyc} + 2,456,803$~d.

\textbf{Star \#992} is another very red variable,  $\langle V \rangle - \langle I \rangle = 4.56$ mag, with a moderate photometric range of $R_{\rm V} = 0.71$ mag.  Period analysis on the $V$-band data showed evidence of periodicity at 38.1, 104, 51.5, and 685 days (the middle two are suspiciously close to a 2:1 ratio, and the latter may be an LSP), though the power spectra included many peaks only slightly less strong. We therefore ran the $I$-band data through \textsc{VStar} and obtained strongest power at 53.1, 51.1 and 38.0 days (there is a weaker period at 606 d that may correspond to the longest period from $V$). We therefore advocate periods of 38.1 and 51.3 days, with a possible LSP around 650 days, and suspect that the star is a SRB variable.

\textbf{Variable \#993} is extremely blue,  $\langle V \rangle - \langle I \rangle = 0.80$ mag, and is among the brightest stars in the strip of main sequence stars shown in Figure~1 of the main text. The standard search on the LPV period range of 10-1000~d yielded a strong peak at 15.109~d, which might indicate that \#993 is a Wolf-Rayet or S~Doradus star (it is phased in Figure~\ref{fig_phi} with this period and $E_0 = 2,456,803.7$~d). An optical spectrum is required to confirm this classification. Experience with similar stars in \citet{abbas15} indicates that this star could be a W~Ursae Majoris contact binary, so we tested periods from 0.1-10~d and found a period at 0.93551~d with higher power. Phased light curves with this and the 15-d period appeared similarly good, with the data points scattering evenly around the sine-shaped models. Doubling the shorter period to recreate the behavior of a W~UMa star showed minima with slightly different depths, suggesting that 1.87102~d represents the orbital period of this star (phased with $E_0 = 2,456,802.94$~d in Figure~\ref{fig_phi}), in good agreement with the phased light curves of W~UMa stars shown in \citet{sj96}. However, the period is relatively long for a contact binary, and the absolute magnitude of $M_{\rm V} = -0.7$~mag seems overly luminous (yet underluminous for a WR or SDOR star). Additional data are needed to classify this star with confidence.

\textbf{Variable \#995} has a moderate mean color of $\langle V \rangle - \langle I \rangle = 2.81$ mag and a narrow range of variability, $R_{\rm V} = 0.15$ mag. Analysis with  \textsc{VStar} indicates two long periods of 602 and 284 days which are apparent individually in the star's light curve, and when combined, match nicely most of the star's long timescale excursions. The shorter period of 21.1 days can be seen in a phased light curve, and may represent the star's pulsation period in a low-amplitude, higher harmonic mode typical of OSARG pulsation described by \citet{soszynski13}. We propose that this star is an SRB and the two long periods are LSPs, analogous to the case of star \#244. We estimate $M_{\rm V} = -2.0$~mag for star~\#995.

\textbf{Star \#996} has a red color, $\langle V \rangle - \langle I \rangle = 3.92$~mag, and modest range of  $R_{\rm V} = 0.47$~mag.
 \textsc{VStar} finds two shorter periods, 42.0 and 42.8 days, which are not aliased, and which, when combined, interfere to produce a slowly-varying amplitude that approximates that seen in the observed light curve. Adding the longer period at 310 days improves the light curve model even more. We thus suggest that star \#996 is a double-mode semi-regular (SRB) variable with an LSP. This star was  also detected by \citet{jayasinghe18} as  ASASSN-V J190742.80+000440.5 and classified as an SR-type LPV with a period of 29.23~d, mean $V = 15.19$~mag, and amplitude $\Delta V = 0.28$~mag. Our \textsc{VStar} analysis found a period of 29.3~d at the eighth-ranked peak in the power spectrum, raising the possibility that pulsation activity shifts energy between different modes over time in this star. 
 As was the case for star \#27, our median magnitude ($V$ = 15.44~mag) is fainter and our range is larger, consistent with blended light affecting the ASAS-SN photometry. 
 We estimate $M_{\rm V} = -1.0$~mag for star~\#996.
 
\textbf{Star \#987:} Like star \#993, our new variable \#987 has a blue color of  $\langle V \rangle - \langle I \rangle = 1.36$~mag that places it on the strip of main sequence stars shown on the CMD in Figure~1 of the main paper. The star has a moderate photometric range of  $R_{\rm V} = 0.72$ mag, much of which is explained by the strong period found at 0.18296~d. The resulting phased light curve is quite sine-shaped, and doubling the period gives a light curve with eclipses of similar depth typical of a W~Ursae Majoris contact binary with a period of 0.36592~d, as shown in Figure~\ref{fig_phi} (where it is phased with $E_0 = 2,456,803.425$~d). This classification is consistent with our estimate of $M_{\rm V} = +3.3$~mag.
 
\textbf{Star \#988:} This variable is red, $\langle V \rangle - \langle I \rangle = 3.70$~mag, and of moderate range,  $R_{\rm V} = 0.49$~mag. The first three periods from  \textsc{VStar} (41.8, 32.9, and 104 days), when combined, provide a model that explains many of the features in the time-magnitude plot. A longer period of 699~d with a similar amplitude improves the match between the model and observed data. We classify this star as a multii-periodic semi-regular (SRB) with a possible LSP.

\section{Variables detected by the wide field imagers} \label{sec_wide}

Though the lower spatial resolution and shorter time span of the images in the P1AU data set make it less-suitable for detecting variable stars than the P5AA set, its field of view covers more than five times the area so it may contain additional variable stars.  The time span and observing cadence make it difficult to determine periods of variables with periods less than about 30 days -- these stars' time-magnitude plots would show scatter which could not be corroborated as real by the process of finding the period and producing a phased light curve showing clear cyclic behavior as was done for the stars \#993 and \#987 in the previous section.  Therefore, we focus our attention on the detection of stars that show slow, correlated variations in $V$ and $I$ in their time-magnitude plots, as well as large values of the \citet{welch-stetson93} variability index and large light variations.  We acknowledge the resulting bias towards finding long period variable stars, as well as our prioritization of red color and uncrowded environment. 

Because the BG and P1 data sets alone provided far fewer points and less time coverage, in most cases we only present the single best period from the \textsc{VStar} analysis. The long observational gap between the BG and P1 data sets ($\sim$1400 days) created a ``picket fence'' of closely-spaced periods in the power spectrum with the interval between periods related to the addition or removal of a periodic cycle of the star during the gap. We selected the period with the highest power, but acknowledge that this choice may be influenced by unidentified aliases, and that more data is required to obtain definitive periods for the following stars.

\textbf{Star \#085} was discovered in the BGSU data presented in \citet{wehrung13} and is listed in the VSX as ``[WLR2013]~85.'' Here we add the star's data from the P1AU set and determine its mean color to be $\langle V \rangle - \langle I \rangle = 4.10$~mag and its photometric range to be $R_{\rm V} = 0.40$~mag. Using  \textsc{VStar} we find the period with the most power to be 42~d with a sinusoidal amplitude of 0.090~mag. These properties are consistent with star \#085 being a semi-regular (SR) long period variable.

\textbf{Variable \#132}, or  ``[WLR2013]~132,'' is a very red star with $\langle V \rangle - \langle I \rangle = 4.50$~mag and a range of $R_{\rm V} = 0.42$~mag. A single, clear period at 43~days indicates it is probably a semi-regular LPV (SR).

\textbf{Star \#419}, or ``[WLR2013]~419,'' is also quite red, with $\langle V \rangle - \langle I \rangle = 4.29$~mag and a large range of at least $R_{\rm V} = 1.83$~mag that is best seen in the new P1AU data as a slow decline, minimum and rise spanning less than one full light cycle. The best period is 185~d, which, by itself, explains most of the variation in the available data. When several cycles of this stars pulsation are well mapped, we suspect the amplitude may be near or above the $R_{\rm V} = 2.5$~mag limit used to separate semi-regular from Mira long period variables. Until then, we classify it as a SRA variable.

\textbf{Star \#105:} This newly-detected variable star has a red color of $\langle V \rangle - \langle I \rangle = 3.43$~mag  and a range of  $R_{\rm V} = 0.27$~mag.  A period of 24~days ($A = 0.046$~mag) and one with 139~d ($A = 0.054$~mag)  had similar power, and could represent a pulsation period and LSP in this SR variable. This star was also detected by  \citet{jayasinghe18} as ASASSN-V J190813.72+000019.4 with mean $V$ = 14.77~mag, $V$ amplitude 0.27~mag and an irregular ``L'' classification with no discernible period. The ASAS-SN photometry agrees well with ours for this star, which has no bright stars nearby on our images that might compromise photometry. We estimate $M_{\rm V} = -1.3$~mag.

\textbf{Star \#154} is another new variable detected in the P1AU data set whose red color of $\langle V \rangle - \langle I \rangle = 4.21$~mag  and  range of   $R_{\rm V} = 0.48$~mag suggest it is another semi-regular LPV. Our \textsc{VStar} analysis support this interpretation with strong power at periods of 43~d and 73~d that do a good job of matching the observed light curve, leading us to think the star is a multi-periodic pulsator (SRB). We estimate $M_{\rm V} = -0.1$~mag.

\textbf{Star \#184:} This new variable is red ($\langle V \rangle - \langle I \rangle = 3.42$~mag) with $R_{\rm V} = 0.36$~mag and a single period of 54~d. It is probably a SR variable.

\textbf{Star \#270} is the same as ASASSN-V J190726.81-000347.7 which \citet{jayasinghe18} found to have mean $V = 15.09$ mag, amplitude $\Delta V = 1.15$ mag, period 79.51~d, and to be a semi-regular variable. In our data, \#270 is red ($\langle V \rangle - \langle I \rangle = 3.37$~mag     but has a larger range of $R_{\rm V} = 1.97$~mag and dimmer mean magnitude (15.18~mag) than the ASAS-SN results, suggesting that blending with nearby stars of constant brightness may have diluted the ASAS-SN light curve somewhat. Our  \textsc{VStar} period of 79.6~d is very strong and is in excellent agreement with that of ASAS-SN. We classify \#270 as a SRA-type LPV but note that its strong, high-amplitude variation is near the threshold to be considered a Mira. We estimate $M_{\rm V} = -0.9$~mag for star~\#270.

\textbf{Variable \#358} is quite red, with $\langle V \rangle - \langle I \rangle = 4.69$~mag,   and  $R_{\rm V} = 0.52$~mag. The strongest power is at a period of 243~d, which describes the slower variation in the P1AU data, and the addition of a weaker period of 60~d accounts for much of the short-timescale variations. We tentatively attribute them to an LSP and the pulsation period, respectively, and classify the star as a semi-regular variable (SR).

\textbf{Star \#451} has an extremely red color of $\langle V \rangle - \langle I \rangle = 5.49$~mag  and a range that is likely larger than the value  $R_{\rm V} = 1.15$~mag we report. The BGSU data are noisy and do little to constrain the star's period, while during the P1AU data set the star rose steadily from $V \approx 17.75$~mag and past a peak at $V \approx 16.95$~mag, covering less than a full cycle (see Figure~5 of the main paper). \textsc{VStar} provides several periods with similar power (733, 489, 368, 295, and 250~d in declining order of power and amplitude) and we cannot distinguish which is correct. We suspect the star may be a Mira LPV, but until more photometry is available to refine the period and amplitude, we provisionally classify \#451 as a SRA variable.

~
~
~

\vspace{5mm}
\facilities{AAVSO, CTIO:PROMPT}

\software{IRAF (\url{https://iraf.net) }
\software{\textsc{VStar} (https://www.aavso.org/vstar) }
          }